\documentclass[11pt]{article}

\usepackage{ACL2023}

\usepackage{times}
\usepackage{latexsym}
\usepackage{amsthm}
\usepackage{amsmath}
\usepackage{amssymb}
\usepackage{xcolor}
\usepackage{graphicx}
\usepackage{booktabs}
\usepackage[table]{xcolor}
\usepackage[T1]{fontenc}
\usepackage{multirow}
\usepackage{subcaption}
\usepackage{adjustbox}
\usepackage{tabularx}
\usepackage[utf8]{inputenc}

\usepackage{bm}
\usepackage{soul}
\usepackage{caption}
\usepackage[linesnumbered,ruled,vlined]{algorithm2e}
\usepackage{algpseudocode}
\usepackage{hyperref}

\usepackage{etoolbox}
\AtBeginEnvironment{equation}{\vspace{-0.5em}}
\usepackage{microtype}

\usepackage{inconsolata}

\usepackage{enumitem}
\usepackage{tikz}
\usepackage{xcolor}

\def\mystrut(#1,#2){\vrule height #1pt depth #2pt width 0pt} 

\definecolor{myred}{rgb}{1.0, 0.8, 0.8}   % Light red
\definecolor{mygreen}{rgb}{0.8, 1.0, 0.8} % Light green
\definecolor{myblue}{rgb}{0.8, 0.8, 1}

\newcommand{\gcell}[2]{%
  \ifdim #1 pt > 0 pt
    \cellcolor{mygreen}#2%
  \else
    \cellcolor{myred}#2%
  \fi
}

\newcommand{\bcell}[2]{%
  \ifdim #1 pt > 0 pt
    \cellcolor{myblue}#2%
  \else
    \cellcolor{myred}#2%
  \fi
}

\newcommand{\rcell}[2]{%
  \ifdim #1 pt > 0 pt
    \cellcolor{mygreen}#2%
  \else
    \cellcolor{myred}#2%
  \fi
}

\title{\textit{More Than Efficiency:} \\ 
Embedding Compression Improves Domain Adaptation in Dense Retrieval}

\author{Chunsheng Zuo \\
  Johns Hopkins University \\
  \texttt{czuo3@jh.edu} \\\And
  Daniel Khashabi \\
  Johns Hopkins University\\
  \texttt{danielk@jhu.edu} \\}

\definecolor{darkgreen}{RGB}{0,140,0}

\begin{document}
\maketitle
\begin{abstract}
Dense retrievers powered by pretrained embeddings are widely used for document retrieval but struggle in specialized domains due to the mismatches between the training and target domain distributions. Domain adaptation typically requires costly annotation and retraining of query-document pairs. In this work, we revisit an overlooked alternative: applying PCA to domain embeddings to derive lower-dimensional representations that preserve domain-relevant features while discarding non-discriminative components. Though traditionally used for efficiency, we demonstrate that this simple embedding compression can effectively improve retrieval performance. Evaluated across 9 retrievers and 14 MTEB datasets, PCA applied solely to query embeddings improves NDCG@10 in 75.4\% of model-dataset pairs, offering a simple and lightweight method for domain adaptation.
\end{abstract}

\section{Introduction}

With recent advancements in retrieval-augmented generation~\cite{lewis-rag-2020, fan-ragsurvey-2024}, dense retrievers~\cite{karpukhin-etal-2020-dense} have become increasingly prominent. These models produce vector embeddings that encode semantic information from text, enabling effective matching and retrieval of contextually relevant documents. However, dense retrievers pretrained on general, large-scale corpora inherently encode semantic features reflecting the training data distribution~\cite{reimers-gurevych-2019-sentence}. Consequently, when applied to specialized domains such as biomedical or finance, these pretrained retrievers face challenges due to distribution shifts between training data and domain-specific text~\cite{lupart-msmacrodistributionshift-2023}. Such shifts can lead retrievers to overlook critical domain-specific information or fail to capture essential nuances during the retrieval process.

\begin{figure}[ht]
    \centering
    \vspace{-0.4cm}
    \includegraphics[width=\linewidth, trim=0.5cm 0.5cm 0.5cm 0.2cm, clip]{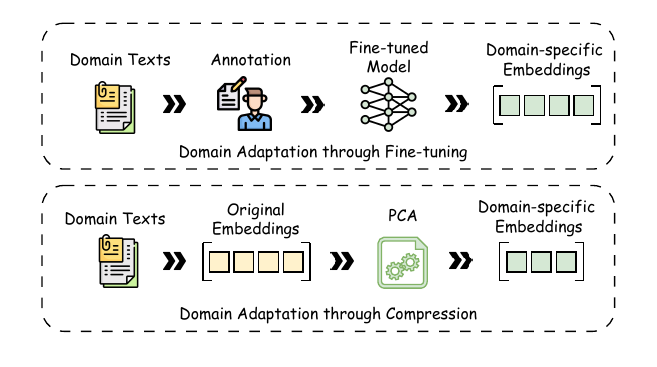}
    \vspace{-0.65cm}
\caption{PCA compresses embeddings to become domain-specific, providing a more efficient domain adaptation method than traditional fine-tuning, with an additional advantage of a lower retrieval cost. }
    \label{fig:title-image}
\end{figure}

Recent works~\cite{gururangan-etal-2020-dont,siriwardhana2023improving, domaindescription2023,li-gaussier-2024-domain} have addressed domain adaptation by fine-tuning retrievers on annotated domain-specific datasets, typically comprising query-document pairs. Nonetheless, acquiring high-quality domain-specific annotated data can often be tedious due to annotation costs and data availability constraints. 

In this work, we explore a simpler yet overlooked approach for domain adaptation: dimensionality reduction through Principal Component Analysis (PCA) \cite{abdi2010principal}. While traditionally employed for dimensionality reduction to speed up retrieval \cite{ma-etal-2021-simple}, we hypothesize that PCA also highlights principal dimensions critical to the domain’s semantic space, refocusing retrieval on domain-relevant information without introducing new knowledge. We posit that PCA, by identifying principal components in a domain-specific corpus or query embeddings, implicitly highlights dimensions critical to the domain's semantic space. Retaining only the top principal dimensions thus refocuses the retriever's attention on domain-relevant information.

Our contribution includes:
\begin{itemize}
\item We demonstrate that a simple application of PCA to queries alone improves retrieval performance (NDCG@10) in a majority (75.4\%) of the evaluated model-dataset combinations.
\item To the best of our knowledge, we are the first to establish that fitting PCA solely on queries is more effective for domain adaptation than using both queries and documents.
\item We investigate the relationship between model domain familiarity and performance gains from PCA, revealing divergent, model-specific adaptation patterns.
\item  We show that PCA, a zero-cost method, can outperform a state-of-the-art compressive domain adaptation technique~\cite{thakur2023injectingdomainadaptationlearningtohash} that relies on computationally expensive pseudo-label curation and fine-tuning.

\end{itemize}

\section{Related Works}
\paragraph{Unsupervised Domain Adaptation with Synthetic Data:}  
Recent works have explored domain adaptation through synthetic data generation and pseudo-labeling. \citet{wang-etal-2021-tsdae-using, wang-etal-2022-gpl} leverage denoising auto-encoders and pseudo-relevance labeling to adapt retrievers without annotated data, while \citet{domaindescription2023, meng2022augtriever, InPars2022} generate synthetic queries and documents through text generation models. 
Building on Generative Pseudo Labeling (GPL)~\cite{wang-etal-2022-gpl}, the work \citet{thakur2023injectingdomainadaptationlearningtohash} apply Joint Product Quantization~\cite{JPQ} to a GPL‑trained Tas‑B model~\cite{tasb} to enable compact, efficient retrieval at a modest effectiveness cost. These approaches demonstrate that synthetic data can mitigate domain shift, but they require substantial computation for data generation and fine‑tuning. In contrast, our study isolates the underexplored role of PCA as a simple, training‑free compression/adaptation technique. 
% Although we do not compare directly to \citet{thakur2023injectingdomainadaptationlearningtohash}, we include a short comparison result in Appendix~\ref{App:ida-comparison}.
\paragraph{Few-Shot and Zero-Shot Adaptation:}
Alternative approaches minimize annotation requirements by using large language models (LLMs) for few-shot data generation. \citet{dai2022promptagator} create task-specific retrievers from as few as 8 examples via LLM-generated queries, and \citet{huang-etal-2023-converser} adapt conversational retrievers using only 6 dialogue examples. While effective, these methods depend on the availability of powerful LLMs and carefully crafted prompts. Additionally, Promptriever~\cite{weller2024promptriever} can adapt to a new domain by following task-related instructions. 
% Methods such as \citet{morris2024contextual} adapt retrievers by structurally modifying their architecture and retraining from scratch using unannotated domain-specific corpora.
Methods such as \citet{morris2024contextual} adapt retrievers by modifying the architecture and training objective to incorporate neighboring-document information, improving out-of-domain retrieval.

Despite avoiding reliance on annotations, these methods require substantial changes to the model's architecture or training process, inapplicable to other already-trained models.

\paragraph{PCA for Retrieval:}  
Prior work has used dimensionality reduction primarily to improve retrieval efficiency \cite{ma-etal-2021-simple}. \citet{raunak-etal-2019-effective} apply PCA to pre-trained embeddings for downstream task performance, while \citet{chen2020image} binarizes PCA-compressed embeddings for image retrieval. However, these methods either optimize for training data characteristics or sacrifice embedding expressivity, limiting their adaptability.

\section{Preliminaries and Notation}

We set up the framework and notations, starting with defining the retrieval problem (\S\ref{subsec:notation}), followed by matrix projection (\S\ref{subsec:pca}).

\subsection{Standard Dense Retrieval}
\label{subsec:notation}

\paragraph{The retrieval problem:} 
Given a collection of documents $D=\{d_j\}_{j=1}^{m}$ and a set of likely queries $Q=\{q_i\}_{i=1}^{n}$, the fundamental task is to, given query $q_i \in Q$, 
retrieve a ranked list of the most relevant documents in $D$: 

\begin{equation}
\label{eq:1}
\text{top-}k_{j \in [1, m]}(\text{sim}(q_i, d_j)),
\end{equation}
where $\text{sim}(., .)$ 
is a similarity metric that quantifies relevance between a query $q_i$ and a document $d_j$. 

\begin{figure*}[th]
    \centering
    \includegraphics[width=\linewidth, trim=0cm 4cm 0.0cm 3cm, clip]{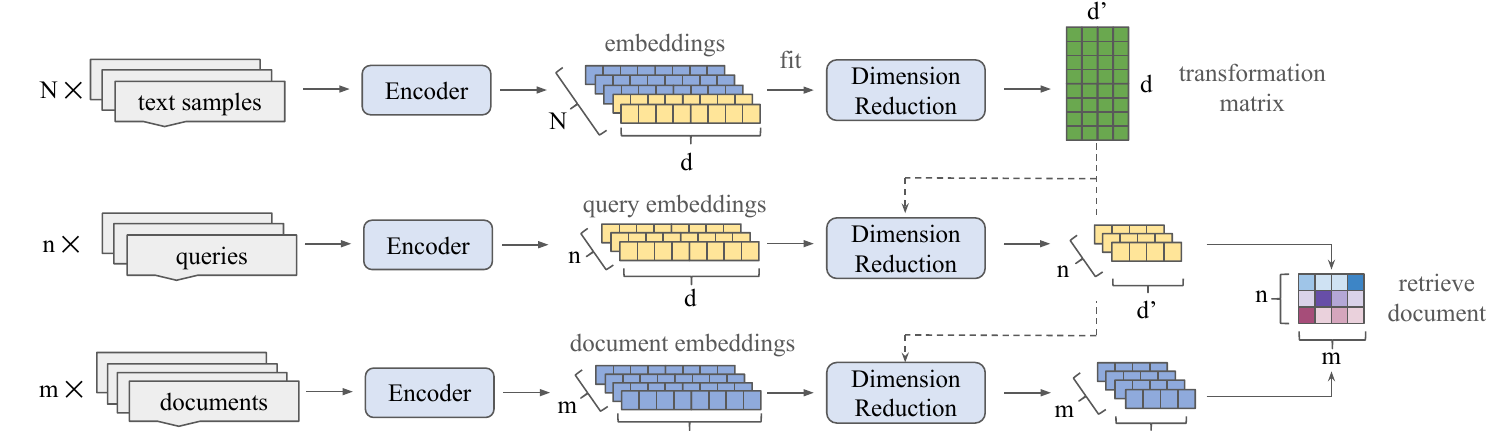}
    \caption{Pipeline for adapting embeddings to a test time domain using PCA (\S\ref{subsec:pca}). The encoder in the top branch processes text samples (queries, documents, or a mixture) from the test domain. These samples are used by the PCA algorithm to learn a projection matrix of the top $d'$ principal components. This matrix then transforms both the test domain's query and document embeddings. Finally, a similarity match is performed on the transformed embeddings to retrieve the most relevant documents.}
    \label{fig:pipeline}
\end{figure*}

\paragraph{Dense retrieval:}
In the special case of dense retrieval, a pre-trained neural network, which we denote as the \text{ENC}, maps both queries and documents into shared high-dimensional embeddings: 
\[
\mathbf{q}_i \leftarrow \text{ENC}(q_i)\in\mathbb{R}^{d}, \;
\mathbf{d}_j \leftarrow \text{ENC}(d_j)\in\mathbb{R}^{d}.
\]
Given these dense representations, the similarity between a query $q_i$ and a document $d_j$ is computed via the cosine similarity between their $d$-dimensional representation: 
\begin{equation}
\label{eq:2}
\text{sim}(q_i,d_j):=
\frac{\mathbf{q}_i\!\cdot\!\mathbf{d}_j}
     {\lVert\mathbf{q}_i\rVert\,\lVert\mathbf{d}_j\rVert},
\end{equation}
which is used to rank and retrieve documents for each query based on this similarity score. 
Our work builds upon this foundation by introducing a dimensionality reduction step for domain adaptation after the initial encoding.

\subsection{Matrix Projection with PCA}
\label{subsec:pca}

PCA is a widely used framework for 
projecting high-dimensional data to a lower-dimensional space.
Let $\mathbf{X} \in \mathbb{R}^{k \times d}$ represent the matrix of $k$ items (e.g., documents), each with $d$-dimensional embeddings. The first step is mean-centering.
Let $\tilde{\mathbf{X}} = \mathbf{X} - \mathbf{1}_k\boldsymbol{\mu}^\top$ be the mean-centered version, where $\boldsymbol{\mu} = \frac{1}{k}\mathbf{X}^\top\mathbf{1}_k$ is the column-wise mean vector and $\mathbf{1}_k$ is a $k$-dimensional vector of ones.

To reduce dimensionality from $\mathbb{R}^{k \times d}$ to $\mathbb{R}^{k \times d'}$, PCA finds an orthogonal projection matrix $\mathbf{W} \in \mathbb{R}^{d \times d'}$ that maximizes variance in the projected data:
\begin{equation}
\mathbf{W}^* = \text{argmax}_{\mathbf{W}^\top \mathbf{W}=I_{d'}} \mathbf{W}^\top \tilde{\mathbf{X}}^\top \tilde{\mathbf{X}} \mathbf{W},
\end{equation}
where $I_{d'}$ is the $d' \times d'$ identity matrix.
\paragraph{Solving this objective:}
The solution $\mathbf{W}^*$ consists of the top $d'$ \textit{eigenvectors} of the sample 
covariance matrix $\frac{1}{k-1}\tilde{\mathbf{X}}^\top \tilde{\mathbf{X}}$, corresponding to the directions of greatest variance. The associated \textit{eigenvalues} quantify how much variance each principal axis explains.
This solution can also be obtained directly via the Singular Value Decomposition (SVD) of the mean-centered matrix $\tilde{\mathbf{X}}$.

\section{Method: Unsupervised Representation Compression for Retrieval}

We detail our approach for adapting pre-trained dense retrievers to a target domain using dimensionality reduction. 
% We first define the standard retrieval process and notation (\S\ref{subsec:notation}), then introduce our method (\S\ref{subsec:pca}) and finally discuss the theoretical motivation behind it (\S\ref{subsec:theory}).

% \subsection{Domain Adaptation via PCA Projection}
% \subsection{Retrieval via Projected Embeddings}
\label{subsec:ours}

\paragraph{Retrieval on down-projected embeddings:}
The goal is to project the high-dimensional embeddings into a lower-dimensional subspace $\mathbb{R}^{d'}$ (where $d' < d$) that effectively captures the semantic structure of the target domain, by discarding information that does not allow distinguishing the target domain. Formally, 
let $\mathbf{W}$ be a projection matrix used to linearly transform both query and document embeddings into the lower-dimensional space:
\[
\mathbf{q}'_i \leftarrow \mathbf{q}_i \mathbf{W}, \qquad
\mathbf{d}'_j \leftarrow \mathbf{d}_j \mathbf{W}.
\]
These down-projections effectively re-align the original embedding features, emphasizing semantic components aligned with the target domain. 
Note that none of this involves adapting the parameters of the pre-trained \texttt{Encoder}.

Using the down-projected embeddings, retrieval is carried out in the reduced space as defined in Eq.~\ref{eq:1} and~\ref{eq:2}. We now describe how the projection matrix $\mathbf{W}$ is derived.

\newcommand{\subscript}[1]{$#1$}

\paragraph{Down-projection with PCA:}
At the core of our approach is learning a projection matrix $\mathbf{W} \in \mathbb{R}^{d \times d'}$ that reduces the dimensionality of embeddings obtained from the target domain. This matrix is obtained by applying Principal Component Analysis (PCA) to embeddings produced by a fixed, pre-trained dense retriever (see Fig.~\ref{fig:pipeline}).
Specifically, we explore two strategies for fitting the PCA:

\begin{enumerate}[leftmargin=*, topsep=1pt,itemsep=0.1pt,partopsep=1pt,label=(\subscript{{\arabic*}})]

\item 
\underline{\emph{Query Compression:}} The PCA model is computed exclusively on the set of query embeddings $\{\mathbf{q}_i\}_{i=1}^{n}$. This approach aims to find a subspace that maximizes the variance observed within the query distribution, potentially highlighting dimensions crucial for distinguishing query intents in the target domain.

\item 
\underline{\emph{Query+Document Compression:}} The PCA model is computed on the union of query and document embeddings, $\{\mathbf{q}_i\}_{i=1}^{n} \cup \{\mathbf{d}_j\}_{j=1}^{m}$. This captures the variance across the entire target corpus to find a subspace shared by both query and document semantics.
\end{enumerate}
See Alg.~\ref{alg:pca-domain-adapt} for the details of the construction.

\begin{algorithm}[th]
\caption{PCA Domain Adaptation}
\label{alg:pca-domain-adapt}
\small
\DontPrintSemicolon
\KwIn{\texttt{Encoder} (pretrained retriever), query set $Q = \{q_i\}_{i=1}^n$, document set $D = \{d_j\}_{j=1}^m$, target dim $d' < d$}
\KwOut{Ranked document list for each $q_i \in Q$}

Encode queries: $\mathbf{q}_i \gets \texttt{Encoder}(q_i)$ for all $q_i \in Q$\;
Encode docs: $\mathbf{d}_j \gets \texttt{Encoder}(d_j)$ for all $d_j \in D$\;

Fit PCA on $\{\mathbf{q}_i\}$ or $\{\mathbf{q}_i\} \cup \{\mathbf{d}_j\}$ to obtain mean $\boldsymbol\mu$ and projection matrix $\mathbf{W} \in \mathbb{R}^{d \times d'}$\;

Project: $\mathbf{x}' \gets (\mathbf{x} - \boldsymbol\mu)\mathbf{W}$ for all $\mathbf{x} \in \{\mathbf{q}_i\} \cup \{\mathbf{d}_j\}$\;

\ForEach{$q_i \in Q$}{
  \ForEach{$d_j \in D$}{
    Compute similarity: $s_{ij} \gets \frac{{\mathbf{q}_i'} \cdot {\mathbf{d}_j'}}{\|\mathbf{q}_i'\| \|\mathbf{d}_j'\|}$\;
  }
  Rank $D$ by $s_{ij}$ in descending order\; 
}
\Return ranked document lists for all $q_i$\;
\end{algorithm}

\paragraph{Assumptions:}
Our approach assumes access to a corpus of documents and queries from the target domain.  Importantly, it operates solely on raw queries and documents, \textit{without} relying on labeled query-document relevance pairs typically required for supervised retrieval. This is a practical assumption in many settings where unlabeled data is abundant.

\subsection{Insights Behind Embedding Projection}
\label{subsec:theory}

\begin{figure}[ht]
    \centering
    \includegraphics[width=\linewidth,trim=0cm 0.5cm 0cm 0.7cm]{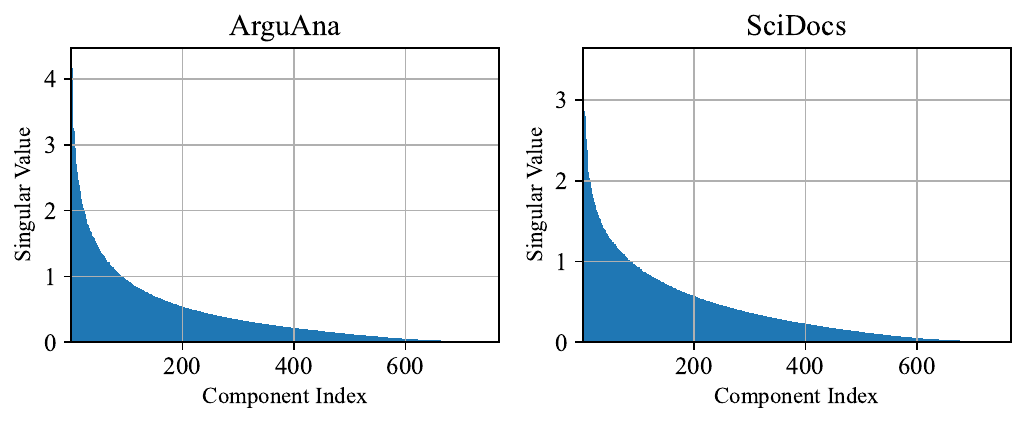}
    \caption{The distribution of the eigenvalues for the principal components after applying PCA to fit the queries' embeddings of the Sentence-T5 model on each dataset (ArguAna and SciDocs).}
    \label{fig:components}
\end{figure}

\paragraph{IR representations are low-rank:}
As noted in \S\ref{subsec:pca}, PCA identifies a low-dimensional subspace whose eigenvectors capture the directions of greatest variance in the target-domain embeddings. The corresponding eigenvalue for each principal axis represents its importance. We noticed that, similar to the examples shown in Fig.~\ref{fig:components}, the eigenvalue spectra for the embeddings of a dataset typically exhibit a power-law-like decay (See Appendix~\ref{App:decay}), facilitating lower-dimensional representation.

\paragraph{The projection helps domain adaptation by preserving salient information for the target:}
The core hypothesis is that these high-variance directions preferentially capture the most salient semantic variations specific to the \textit{target} domain. 
Projecting onto the subspace defined by these directions filters out low-variance components---often associated with \textit{source}-domain artifacts or noise---thereby improving focus on the \textit{target} domain.

% By projecting the embeddings onto the subspace spanned by these principal components and discarding dimensions associated with lower variance, PCA can effectively filter out noise or features that are predominantly relevant to the source domain upon which the encoder was originally trained, thereby enhancing focus on the test domain. 

This approach relies on three key premises:
\begin{enumerate}[itemsep=0pt, parsep=0pt,topsep=0pt,leftmargin=*,]
    \item The pre-trained encoder, even without fine-tuning, captures sufficient domain-relevant information within its embeddings.
    \item The primary semantic variations characteristic of the \textit{target} domain manifest as directions of high variance in the embedding space.
    \item  Variance from the \textit{source} domain or noise often lies orthogonal to the \textit{target} domain’s principal components and is attenuated by projection.
    % Variance stemming from the source domain or irrelevant noise tends to align with directions orthogonal to the principal components of the target domain data and can thus be reduced by the projection.
\end{enumerate}

\section{Experimental Setup}

\begin{table*}[htbp]
\centering
\scriptsize
\renewcommand{\arraystretch}{1.1}
\setlength{\tabcolsep}{1pt}
\begin{tabular}{l ccccccccc ccccccccc}
\toprule
\multirow{2}{*}{\textbf{Datasets}$\downarrow$} & \multicolumn{9}{c}{\textbf{Query Compression}} & \multicolumn{9}{c}{\textbf{Query+Document Compression}} \\
\cmidrule(lr){2-10} \cmidrule(lr){11-19}
 & \textbf{MiniLM} & \textbf{GTE} & \textbf{Instr.} & \textbf{Dis-Ro} & \textbf{MP-QA} & \textbf{MP-All} & \textbf{Sent-T5} & \textbf{BGE} & \textbf{SFR} & \textbf{MiniLM} & \textbf{GTE} & \textbf{Instr.} & \textbf{Dis-Ro} & \textbf{MP-QA} & \textbf{MP-All} & \textbf{Sent-T5} & \textbf{BGE} & \textbf{SFR} \\
\midrule
\textbf{Code}        & \gcell{1.2}{1.2} & \gcell{2.3}{2.3} & \gcell{-1.0}{-1.0} & \gcell{6.0}{6.0} & \gcell{1.1}{1.1} & \gcell{1.3}{1.3} & \gcell{38.7}{38.7} & \gcell{4.0}{4.0} & \gcell{-0.2}{-0.2} & \bcell{-0.7}{-0.7} & \bcell{1.0}{1.0} & \bcell{-1.8}{-1.8} & \bcell{0.7}{0.7} & \bcell{-1.0}{-1.0} & \bcell{-1.3}{-1.3} & \bcell{18.8}{18.8} & \bcell{-1.3}{-1.3} & \bcell{-0.6}{-0.6} \\
\textbf{Apps}        & \gcell{-4.1}{-4.1} & \gcell{1.1}{1.1} & \gcell{-14.5}{-14.5} & \gcell{-16.2}{-16.2} & \gcell{1.1}{1.1} & \gcell{-11.1}{-11.1} & \gcell{18.0}{18.0} & \gcell{-16.1}{-16.1} & \gcell{-0.2}{-0.2} & \bcell{-21.1}{-21.1} & \bcell{-1.5}{-1.5} & \bcell{-24.8}{-24.8} & \bcell{-52.8}{-52.8} & \bcell{-12.4}{-12.4} & \bcell{-22.1}{-22.1} & \bcell{5.1}{5.1} & \bcell{-36.8}{-36.8} & \bcell{-1.2}{-1.2} \\
\textbf{SciDocs}     & \gcell{-0.2}{-0.2} & \gcell{18.5}{18.5} & \gcell{3.0}{3.0} & \gcell{-0.5}{-0.5} & \gcell{-1.0}{-1.0} & \gcell{3.3}{3.3} & \gcell{3.8}{3.8} & \gcell{4.8}{4.8} & \gcell{2.1}{2.1} & \bcell{-0.4}{-0.4} & \bcell{13.3}{13.3} & \bcell{2.5}{2.5} & \bcell{0.5}{0.5} & \bcell{-0.3}{-0.3} & \bcell{2.4}{2.4} & \bcell{-5.2}{-5.2} & \bcell{-7.9}{-7.9} & \bcell{1.9}{1.9} \\
\textbf{MedQA}       & \gcell{2.1}{2.1} & \gcell{2.0}{2.0} & \gcell{2.1}{2.1} & \gcell{2.7}{2.7} & \gcell{2.2}{2.2} & \gcell{1.9}{1.9} & \gcell{6.9}{6.9} & \gcell{0.4}{0.4} & \gcell{0.5}{0.5} & \bcell{1.9}{1.9} & \bcell{1.4}{1.4} & \bcell{1.6}{1.6} & \bcell{2.2}{2.2} & \bcell{1.3}{1.3} & \bcell{0.9}{0.9} & \bcell{5.6}{5.6} & \bcell{-0.4}{-0.4} & \bcell{0.5}{0.5} \\
\textbf{ArguAna}     & \gcell{1.2}{1.2} & \gcell{-2.2}{-2.2} & \gcell{3.0}{3.0} & \gcell{0.6}{0.6} & \gcell{0.3}{0.3} & \gcell{-0.3}{-0.3} & \gcell{11.1}{11.1} & \gcell{2.2}{2.2} & \gcell{-1.3}{-1.3} & \bcell{0.4}{0.4} & \bcell{-2.5}{-2.5} & \bcell{1.6}{1.6} & \bcell{-0.4}{-0.4} & \bcell{-1.1}{-1.1} & \bcell{-1.4}{-1.4} & \bcell{7.1}{7.1} & \bcell{0.4}{0.4} & \bcell{-2.0}{-2.0} \\
\textbf{StackOverflow} & \gcell{0.1}{0.1} & \gcell{-0.1}{-0.1} & \gcell{-1.1}{-1.1} & \gcell{0.4}{0.4} & \gcell{-0.1}{-0.1} & \gcell{1.2}{1.2} & \gcell{2.9}{2.9} & \gcell{0.4}{0.4} & \gcell{-0.9}{-0.9} & \bcell{0.1}{0.1} & \bcell{0.0}{0.0} & \bcell{-1.1}{-1.1} & \bcell{-0.8}{-0.8} & \bcell{-0.1}{-0.1} & \bcell{0.4}{0.4} & \bcell{2.9}{2.9} & \bcell{-0.1}{-0.1} & \bcell{-1.1}{-1.1} \\
\textbf{TV2Nord}    & \gcell{40.2}{40.2} & \gcell{4.4}{4.4} & \gcell{2.8}{2.8} & \gcell{17.1}{17.1} & \gcell{17.7}{17.7} & \gcell{7.4}{7.4} & \gcell{1.2}{1.2} & \gcell{0.3}{0.3} & \gcell{5.5}{5.5} & \bcell{24.9}{24.9} & \bcell{2.7}{2.7} & \bcell{2.6}{2.6} & \bcell{11.7}{11.7} & \bcell{4.2}{4.2} & \bcell{4.0}{4.0} & \bcell{-0.6}{-0.6} & \bcell{-0.2}{-0.2} & \bcell{2.7}{2.7} \\
\textbf{GerDa}       & \gcell{6.2}{6.2} & \gcell{3.5}{3.5} & \gcell{15.7}{15.7} & \gcell{-1.4}{-1.4} & \gcell{7.4}{7.4} & \gcell{7.0}{7.0} & \gcell{3.2}{3.2} & \gcell{1.8}{1.8} & \gcell{7.8}{7.8} & \bcell{-1.7}{-1.7} & \bcell{-10.1}{-10.1} & \bcell{2.0}{2.0} & \bcell{-10.8}{-10.8} & \bcell{-1.6}{-1.6} & \bcell{-0.6}{-0.6} & \bcell{-7.9}{-7.9} & \bcell{-14.2}{-14.2} & \bcell{-13.4}{-13.4} \\
\textbf{ARC}         & \gcell{2.5}{2.5} & \gcell{1.4}{1.4} & \gcell{12.9}{12.9} & \gcell{2.7}{2.7} & \gcell{3.0}{3.0} & \gcell{2.0}{2.0} & \gcell{11.1}{11.1} & \gcell{7.3}{7.3} & \gcell{4.4}{4.4} & \bcell{3.7}{3.7} & \bcell{1.7}{1.7} & \bcell{13.1}{13.1} & \bcell{2.3}{2.3} & \bcell{3.8}{3.8} & \bcell{1.9}{1.9} & \bcell{9.4}{9.4} & \bcell{8.2}{8.2} & \bcell{3.0}{3.0} \\
\textbf{FeedbackQA}  & \gcell{-1.9}{-1.9} & \gcell{1.0}{1.0} & \gcell{-3.7}{-3.7} & \gcell{0.2}{0.2} & \gcell{-2.2}{-2.2} & \gcell{-3.2}{-3.2} & \gcell{-2.4}{-2.4} & \gcell{-0.4}{-0.4} & \gcell{-0.2}{-0.2} & \bcell{-2.7}{-2.7} & \bcell{0.5}{0.5} & \bcell{-6.0}{-6.0} & \bcell{-6.9}{-6.9} & \bcell{-5.3}{-5.3} & \bcell{-3.5}{-3.5} & \bcell{-4.6}{-4.6} & \bcell{-1.2}{-1.2} & \bcell{-1.2}{-1.2} \\
\textbf{FaithDial}   & \gcell{0.3}{0.3} & \gcell{5.1}{5.1} & \gcell{4.0}{4.0} & \gcell{0.8}{0.8} & \gcell{0.8}{0.8} & \gcell{1.8}{1.8} & \gcell{4.2}{4.2} & \gcell{7.9}{7.9} & \gcell{4.8}{4.8} & \bcell{1.3}{1.3} & \bcell{4.8}{4.8} & \bcell{2.3}{2.3} & \bcell{-0.2}{-0.2} & \bcell{1.3}{1.3} & \bcell{0.8}{0.8} & \bcell{3.1}{3.1} & \bcell{7.9}{7.9} & \bcell{4.6}{4.6} \\
\textbf{MLQA} & \gcell{0.5}{0.5} & \gcell{0.6}{0.6} & \gcell{2.5}{2.5} & \gcell{0.8}{0.8} & \gcell{1.0}{1.0} & \gcell{0.4}{0.4} & \gcell{-0.4}{-0.4} & \gcell{0.2}{0.2} & \gcell{1.8}{1.8} & \bcell{0.2}{0.2} & \bcell{0.6}{0.6} & \bcell{2.5}{2.5} & \bcell{0.4}{0.4} & \bcell{0.8}{0.8} & \bcell{0.3}{0.3} & \bcell{-1.6}{-1.6} & \bcell{0.3}{0.3} & \bcell{1.7}{1.7} \\
\textbf{NarrativeQA} & \gcell{11.0}{11.0} & \gcell{41.5}{41.5} & \gcell{7.1}{7.1} & \gcell{15.3}{15.3} & \gcell{13.1}{13.1} & \gcell{5.8}{5.8} & \gcell{20.5}{20.5} & \gcell{1.6}{1.6} & \gcell{2.6}{2.6} & \bcell{7.1}{7.1} & \bcell{37.0}{37.0} & \bcell{4.7}{4.7} & \bcell{11.6}{11.6} & \bcell{8.5}{8.5} & \bcell{1.5}{1.5} & \bcell{17.4}{17.4} & \bcell{1.1}{1.1} & \bcell{1.6}{1.6} \\
\textbf{SpartQA}     & \gcell{620.0}{620.0} & \gcell{149.5}{149.5} & \gcell{343.0}{343.0} & \gcell{508.4}{508.4} & \gcell{1954.6}{1954.6} & \gcell{511.0}{511.0} & \gcell{119.7}{119.7} & \gcell{192.5}{192.5} & \gcell{849.2}{849.2} & \bcell{626.7}{626.7} & \bcell{173.4}{173.4} & \bcell{717.7}{717.7} & \bcell{463.1}{463.1} & \bcell{1509.1}{1509.1} & \bcell{632.3}{632.3} & \bcell{248.1}{248.1} & \bcell{209.0}{209.0} & \bcell{829.7}{829.7} \\
\midrule
\textbf{Summary} & 11/14 & 12/14 & 10/14 & 10/14 & 11/14 & 10/14 & 12/14 & 10/14 & 9/14 & 9/14 & 8/14 & 8/14 & 7/14 & 8/14 & 8/14 & 9/14 & 6/14 & 8/14 \\
\bottomrule
\end{tabular}
\caption{NDCG@10 improvement (\%) comparison between Query Compression (left) and Query+Document Compression (right). Positive gains are highlighted \colorbox{green!25}{green} for Query Compression and \colorbox{blue!25}{blue} for Query+Document Compression, negative values are \colorbox{red!25}{red}. 
The ``Summary'' row indicates the number of datasets with improvements, per model. 
The table shows that both Query Compression and Query+Document Compression improve retrieval quality. However, \textbf{Query Compression yields more consistent gains, improving 95 out of 126 data–model pairs (75.4\%)}, compared to 71 out of 126 (56.3\%) for Query+Document Compression. More detailed results with raw performance values can be found in Appendix~\ref{app:ndcg10_query_detailed}.
}
\label{tab:merged-compression-results}
\end{table*}

\paragraph{Variations:}
We evaluate three PCA compression strategies.
% For PCA compression, we consider three approaches. 
\textbf{No Compression (Baseline):} Use raw model embeddings without dimensionality reduction.
% \textbf{No Compression (Baseline):} We use the model's raw embeddings without any dimensionality reduction. 
\textbf{Query Compression:} Fit PCA on domain-specific query embeddings and apply the projection to both queries and documents.
% \textbf{Query Compression:} We fit PCA on the embeddings of domain-specific queries only, then apply the resulting transformation to both queries and documents. 
\textbf{Query+Document Compression:} Fit PCA on the combined query and document embeddings to capture broader domain variance.
% \textbf{Query+Document Compression:} We fit PCA on the combined set of query and document embeddings, capturing broader domain variance during the transformation. 
The retention ratio $r$ is defined as:

\begin{equation}
r = \lfloor d'/d \rfloor,
\end{equation}
where $d$ is the original embedding dimension and $d'$ is the number of retained principal components. Unless otherwise noted, we use $r = 0.9$ in our experiments (\S\ref{sec:mainfindings}) and also explore the trade-offs across retention ratios from 0.1 to 1.0 in increments of 0.1.

\paragraph{Datasets:}
We assess PCA-based adaptation on 14 diverse datasets from the MTEB benchmark~\cite{muennighoff-etal-2023-mteb}, comparing against performance with uncompressed embeddings (Appendix~\ref{App:dataset_details}). We also include 11 additional MTEB datasets (Appendix~\ref{app:low}) where the number of queries for each is smaller than the target PCA dimension, leading to lower-than-intended retention ratios.

These datasets span multiple domains (e.g., software engineering, biomedical, finance, and scientific literature, news), and even multiple languages (English, Danish, and German), allowing a comprehensive evaluation of PCA's effectiveness in handling various distribution shifts.

\paragraph{Models:}
We evaluate 9 popular pretrained dense retrieval models with fewer than 2B parameters. Our selection includes five models from the Sentence Transformers library (\texttt{Dis-Ro}, \texttt{MP-QA}, \texttt{MiniLM}, \texttt{MP-All}, \texttt{Sent-T5}) and four other widely used models: \texttt{BGE}, \texttt{GTE}, \texttt{SFR}, and \texttt{Instr}. A complete list of the models, their abbreviations, and corresponding references is provided in Appendix~\ref{app:models}.

\paragraph{Metrics and evaluation:}
We primarily use \texttt{NDCG@10} to evaluate retrieval performance. 
We have also added results for Precision@10 and Recall@10 in Appendix~\ref{App:rec_and_pre}.

Overall, each data-model pair is evaluated under three configurations (baseline, Query Compression, and Query+Document Compression). With 9 models and 14 + 11 datasets, this results in a total of 675 retrieval runs. All experiments are conducted with an RTX 4090 GPU for around 36 hours.

\section{Empirical Results}

\begin{table}[ht]
\centering
\small
\renewcommand{\arraystretch}{1.1}
\setlength{\tabcolsep}{4pt}
\begin{tabular}{lccc}
\toprule
\textbf{Dataset} & \textbf{\# of Queries} & \textbf{\# of Docs} & \textbf{Success Rate} \\
\midrule
MedQA                 & 2048  & 2048    & \cellcolor{mygreen!100}9/9 \\
SpartQA               & 3594  & 1592    & \cellcolor{mygreen!100}9/9 \\
FaithDial             & 2042  & 3539    & \cellcolor{mygreen!100}9/9 \\
NarrativeQA           & 10557 & 355     & \cellcolor{mygreen!100}9/9 \\
ARC                   & 1172  & 9350    & \cellcolor{mygreen!100}9/9 \\
TV2Nord               & 2048  & 2048    & \cellcolor{mygreen!100}9/9 \\
MLQA                  & 11582 & 9916    & \cellcolor{mygreen!88}8/9 \\
GerDa                 & 12234 & 9969    & \cellcolor{mygreen!88}8/9 \\
Code                  & 14918 & 280310  & \cellcolor{mygreen}7/9 \\
ArguAna               & 1406  & 8674    & \cellcolor{mygreen!60}6/9 \\
SciDocs               & 1000  & 25657   & \cellcolor{mygreen!60}6/9 \\
StackOverflow         & 1994  & 19931   & \cellcolor{mygreen!45}5/9 \\
Apps                  & 3765  & 8765    & \cellcolor{mygreen!20}3/9 \\
FeedbackQA            & 1992  & 2364    & \cellcolor{mygreen!10}2/9 \\
\bottomrule
\end{tabular}
\caption{
Per-dataset summary statistics. For each dataset, we report its size (\# of queries and documents) and the success rate of 90\% Query Compression—defined as the proportion of embedder representations that see improved performance after compression.
As shown in the success rate column, \textbf{some datasets consistently benefit from query-only compression across all or many choices of representations}.
% Summarized statistics for dataset sizes and success rate (i.e. the number of models over all models that improve on the dataset) by 90\% Query Compression. \textbf{The success rate is regardless of dataset size.}
}
\label{tab:success_stats}
\end{table}

\subsection{Main Findings}
\label{sec:mainfindings}

\begin{figure*}[h]
    \centering
    \includegraphics[width=\linewidth, trim=0cm 0.4cm 0.0cm 0.6cm, clip=false]{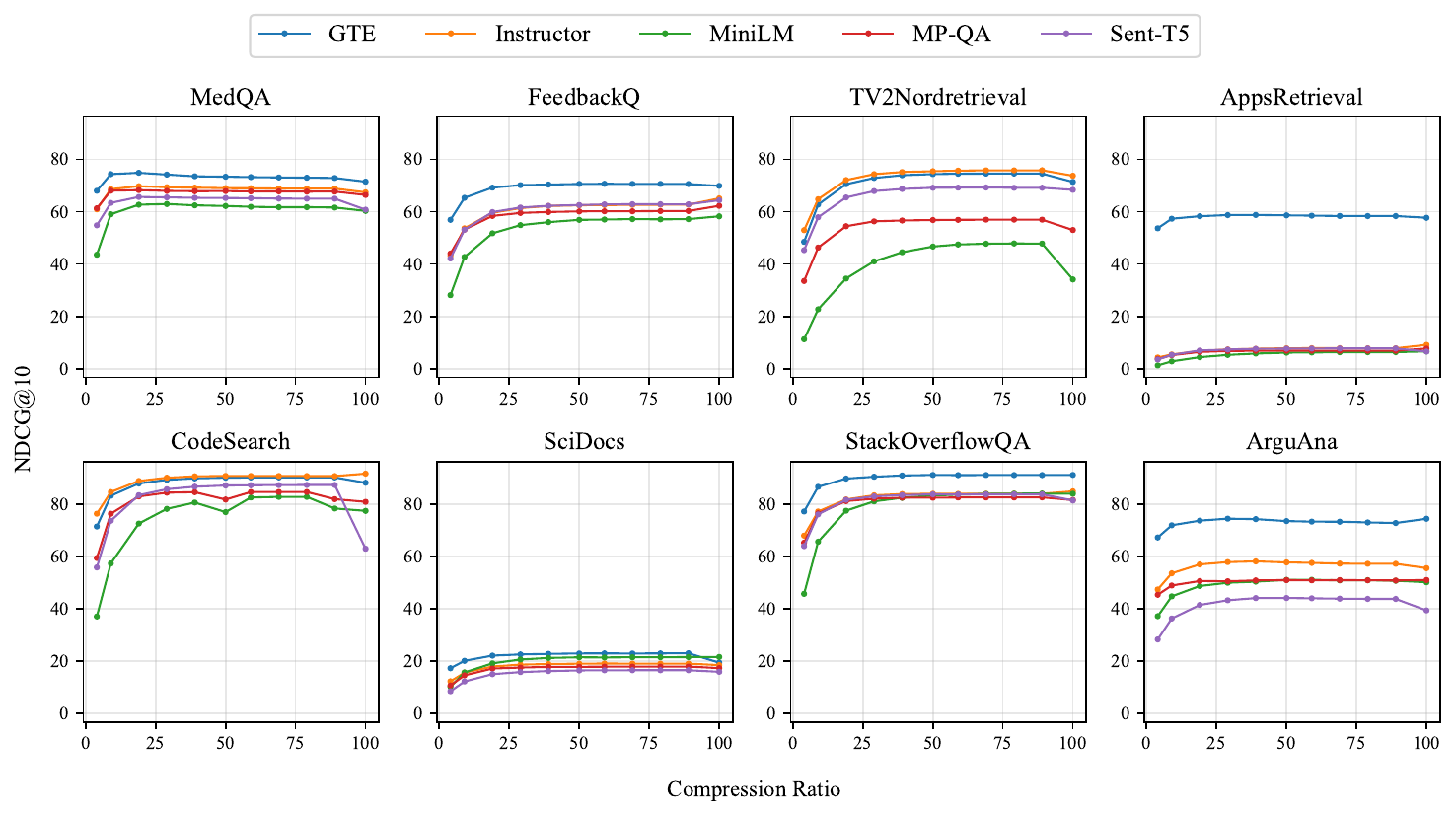}
    % \vspace{-0.1cm}
\caption{NDCG@10 at various retention ratios ranging from 0.1 to 1, with step size of 0.1 and an additional inclusion of a very extreme 0.05. \textbf{While the optimal Query retention ratio is dataset- and model-dependent, in most cases where Query Compression dominates the original embeddings, the dominance can continue even until the retention ratio is around 50\%.}}
    \label{fig:trade-off}
\end{figure*}

\begin{figure*}[t]

    \includegraphics[page=1,width=\linewidth,trim=0cm 1.2cm 0.0cm 0.6cm, clip=false]{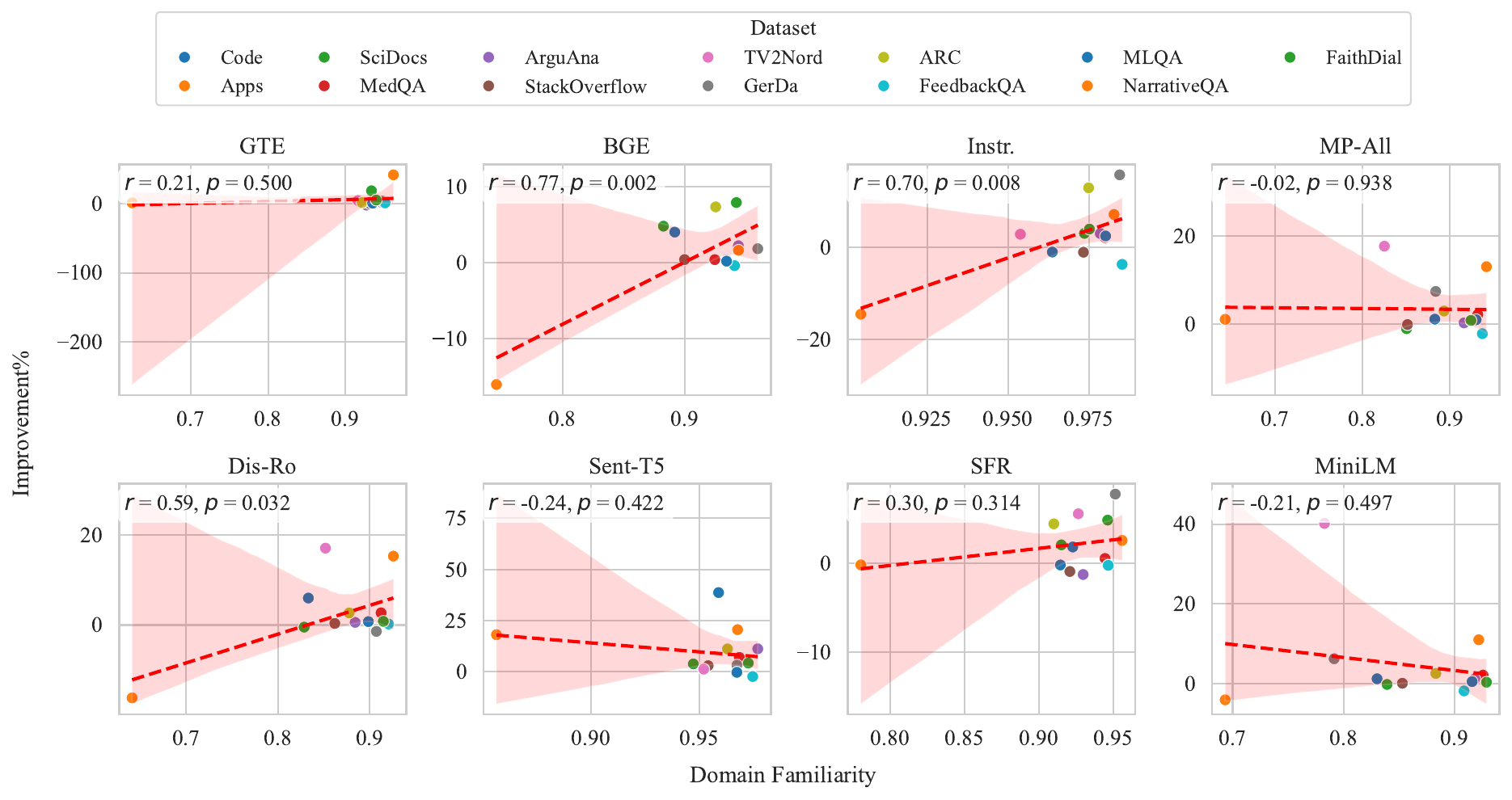}
    \label{fig:sim_sent_t5}
    \caption{Correlation between domain familiarity (Equation~\ref{eq:domainfamiliarity}) and performance improvement after 90\% Query Compression. \textbf{For BGE, Instructor, and Dis-Ro, retrieval performance after compression benefits from higher domain familiarity. }}
    \label{fig:similarity_improvement}
\end{figure*}

\paragraph{Query Compression Generally Improves Performance:}
Table~\ref{tab:merged-compression-results} presents the NDCG@10 improvements obtained from applying 90\% Query Compression across various dataset-model combinations. Query-only compression consistently demonstrates effectiveness across diverse domains, with improvements observed in 75.4\% of model-dataset pairings. All models, except for SFR, achieve improvements in 10 to 11 datasets. Notably, GTE and Sent-T5 show positive gains across 12 out of 14 datasets, making them the most robust models for Query Compression. 
% While most of the degradations are within 4\%, data-model pairs like TV2Nord + MiniLM, NarrativeQA + GTE, and Code + Sent-T5 achieved a huge gain of over 40\%.  
Most degradations are within 4\%. In contrast, several positive outliers are substantial: TV2Nord\,+\,MiniLM, NarrativeQA\,+\,GTE, and Code\,+\,Sent-T5 each achieve gains of over 40\%.

Table~\ref{tab:success_stats} summarizes the success rate of Query Compression on the 14 datasets, where 12 of them have the majority of models improved, and 6 of them have all models improved. Whether a dataset can be successfully adapted through compression is not dependent on the number of queries it contains. Overall, these results highlight that Query Compression generally allows the model to adapt to a new domain with minimal loss or even considerable performance gains.

\paragraph{Query+Document Compression underperforms Query Compression:}

 As a general observation in Table \ref{tab:merged-compression-results}, while Query Compression achieves improvements on 75.4\% of data-model combinations, the percentage drops to 56.3\% for Query+Document Compression. On the Apps dataset, for example, nearly all models experience a substantial performance drop with Query+Document Compression, exemplified by a drastic fall of up to -52.8\% for Dis-RoBERTa. A similar pattern is visible on the GerDa dataset, where Query Compression shows mostly gains, but Query+Document Compression results in significant performance loss for almost every model. This indicates that queries alone are better at capturing domain-specific user information needs, which aligns the PCA projection more precisely with the search task. 
 % In contrast, considering the entire document corpus can dilute this effect. 
 In contrast, high-variance directions in the full document corpus may reflect broad topical or stylistic variation rather than the relevance distinctions expressed by the queries, thereby diluting this effect.
 Despite a few dataset-model combinations that benefit more from Query+Document compression, the overall evidence shows that query-only compression remains the safer, more consistently effective strategy.

\paragraph{Hierarchical Query Structures Amplify PCA Gains:}
While Table~\ref{tab:merged-compression-results} shows that PCA provides widespread benefits, the magnitude of these gains varies significantly across datasets. We argue that this variance is strongly linked to the inherent structure of the domain; specifically, domains with clear hierarchical or categorical query patterns benefit the most from PCA. This effect is exemplified by SpartQA, a dataset organized by a logical structure of attributes like shapes, colors, and spatial relations. Its queries possess a highly regular, almost combinatorial nature, which helps explain the outsized performance improvements observed across all models. Similarly, for MedQA, where models consistently achieve positive gains, the queries are classifiable by a certain medical hierarchy (e.g., diseases, symptoms, treatments). In such structured domains, the principal components identified by PCA are more likely to correspond directly to these meaningful semantic axes. The projection, therefore, acts as a targeted feature selection process, magnifying the benefits of compression.

\subsection{Further Analysis}
\label{sec:further}

\paragraph{Performance at various Retention Ratios:}

We further investigate the trade-off between performance and retention ratio. Figure~\ref{fig:trade-off} shows the NDCG@10 at different Query Retention ratios for 8 of the 14 datasets.

In general, a moderate retention ratio optimizes performance before information loss leads to decline. On CodeSearch, MiniLM peaked at 82.8 NDCG@10 with 80\% of dimensions, up from 77.4, while Sent-T5 saw a dramatic rise from 63.0 to 87.3 with 90\% of dimensions. This suggests an optimal balance between conciseness and informational integrity.

Notably, certain datasets are more tolerant of compression. On ArguAna and MedQA, the performance remains strong down to a 40\% retention ratio. GTE even peaks in MedQA at 10\% compression, equivalent to embeddings with 76 dimensions. We attribute this to the quality of MedQA queries that consist of systematic questions, providing valuable guidance for domain-specific projection. Furthermore, rather than indicating that a few components are essential, being robust to compression could mean that the remaining principal components are sufficiently representative to distinguish between queries/documents within the dataset. 
% Namely, when the retention ratio is smaller than a certain threshold, where not only noise but also certain domain-relevant components are dropped, the remaining components are still discriminative enough for selecting the right documents from the corpus for all queries.

\paragraph{Does PCA Compression Adapt Models Better to Niche Domains than Familiar Ones?}

To investigate whether PCA compression enhances model adaptation to niche domains further, we analyzed the relationship between a model's intrinsic familiarity with a target domain and its post-PCA performance gain. We quantify domain familiarity through \textbf{paraphrasing robustness}, measuring how consistently a model represents semantic equivalence under textual variations. Formally, given a text snippet \( t_i \) and its paraphrases \(\{t^1_i, \dots, t^n_i\}\), an encoder \( E \)'s familiarity to \( t_i \) is defined as:
\vspace{-0.2cm}
\begin{align}
    \text{TF}_n(E, t_i) = \frac{1}{n} \sum_{j=1}^n \text{sim}\left(E(t_i), E(t^j_i)\right),
\end{align}
% \vspace{-0.2cm}
where \(\text{sim}\) denotes cosine similarity. The domain familiarity for dataset \( D = \{t_1, \dots, t_m\} \) with respect to \( E \) is then:
\vspace{-0.2cm}
\begin{align}
    \text{DF}_m(E, D) = \frac{1}{m} \sum_{i=1}^m \text{TF}_n(E, t_i). \label{eq:domainfamiliarity}
\end{align}
% \vspace{-0.2cm}
Higher \(\text{DF}\) indicates stronger familiarity, as the model produces stable embeddings for paraphrased content within the domain. We randomly sampled 10 queries from each dataset, each producing 3 paraphrases (m=10, n=3).\footnote{
Despite the small size, the variance was acceptably low.
% Despite the small sample size, we found the variance sufficiently small.
% Despite the small numbers, we found the variance sufficiently small.
} Fig.~\ref{fig:similarity_improvement} shows the $\text{FD}$ against the post-PCA (90\% compression) performance gain on each dataset \footnote{We do not include SpartQA in this analysis because its performance gain is an outlier.} for 8 of the models.

The results reveal varied trends. Sent-T5, MiniLM, and MP-All exhibit weak negative to negligible correlations, all statistically insignificant. Conversely, Instructor (Instr.) and BGE demonstrate strong, statistically significant positive correlations. Distilled Roberta (Dis-Ro) also shows a statistically significant, moderate positive correlation. Finally, GTE and SFR indicate weak positive correlations that are not statistically significant.

The interpretation of these mixed results remains complex. On one hand, lower domain familiarity (and thus lower $\text{DF}$) might signal an opportunity for PCA to improve embeddings by removing task-irrelevant variance that dominates in unfamiliar contexts; this could align with the negative correlations observed in Sent-T5 and MiniLM. On the other hand, strong positive correlations, as seen with Instructor, BGE, and Distilled Roberta, suggest that PCA might be more effective when the model already has a foundational understanding (higher familiarity) of the domain, possibly by refining already relevant features. The statistically insignificant results for several models, and the presence of both positive and negative trends, suggest that the relationship between domain familiarity and PCA's impact is model-dependent and potentially influenced by other factors not captured by $\text{DF}$ alone.

\subsection{Comparison with Other Domain-Adaptation Methods}
\label{App:ida-comparison}

% Our main text deliberately focuses on the phenomenon of PCA compression—an overlooked yet embarrassingly simple technique that requires no synthetic data generation or model fine-tuning. In that spirit, we avoided mixing PCA with popular domain-adaptation pipelines such as Generative Pseudo-Labeling (GPL) in the core experiments, because doing so would obscure the clarity of our investigation and inflate the computational budget.

% For completeness, however, we report here a side-by-side comparison with Injecting Domain Adaptation (IDA) by \citet{thakur2023injectingdomainadaptationlearningtohash}. IDA augments a GPL-adapted encoder with deep hashing schemes—Joint Product Quantization (JPQ)~\cite{JPQ} and Binary Passage Retrieval (BPR)~\cite{yamada-etal-2021-efficient}—to obtain a compact index. We evaluate on the five datasets used by both our work and IDA. The improvements over the base Tas-B encoder are summarized in Table~\ref{tab:pca-ida}, where numbers for GPL, GPL,+,JPQ, and GPL,+,BPR are copied directly from IDA.

Our main experiments isolate PCA compression as a training-free adaptation mechanism: it requires neither synthetic data generation nor model fine-tuning. This isolation keeps the 9-model, 14-dataset study focused on the effect of the projection itself, rather than on the many design choices introduced by heavier adaptation pipelines. To connect this analysis with stronger adaptation methods, we additionally compare PCA with fine-tuning methods on the subset of datasets shared by both studies.

Specifically, we compare with Injecting Domain Adaptation (IDA) by
\citet{thakur2023injectingdomainadaptationlearningtohash}. IDA augments a
GPL-adapted Tas-B encoder with deep hashing schemes---Joint Product Quantization
(JPQ)~\cite{JPQ} and Binary Passage Retrieval (BPR)~\cite{yamada-etal-2021-efficient}---
to obtain a compact index. We evaluate on the five datasets used by both our work
and IDA. Table~\ref{tab:pca-ida} reports relative improvements over the base Tas-B
encoder; the numbers for GPL, GPL\,+\,JPQ, and GPL\,+\,BPR are taken directly from IDA.

\begin{table}[t]
\centering
\resizebox{\columnwidth}{!}{
\begin{tabular}{lrrrr}
\toprule
Dataset & PCA & GPL & GPL\,+\,JPQ & GPL\,+\,BPR \\
\midrule
SciDocs  & 12.1 & 13.4 &  4.0 &   2.0 \\
ArguAna  & 10.3 & 29.8 &  3.3 & -17.4 \\
FiQA     & 14.3 & 14.6 & -1.6 &   3.0 \\
NFCorpus &  1.9 &  8.1 &  1.5 &  -6.2 \\
SciFact  &  3.2 &  4.8 &  1.8 &  -3.1 \\
\bottomrule
\end{tabular}}
\caption{Relative improvement (\%) over the original Tas-B encoder. PCA matches GPL on SciDocs and FiQA, and outperforms both GPL\,+\,JPQ and GPL\,+\,BPR on all 5 datasets.}
\label{tab:pca-ida}
\end{table}

PCA matches GPL on SciDocs and FiQA, while GPL remains stronger on ArguAna, NFCorpus, and SciFact. PCA also outperforms both compact IDA variants, GPL\,+\,JPQ and GPL\,+\,BPR, on all five shared datasets.  These results suggest that PCA is not a replacement for full synthetic-data fine-tuning when compute is available, but it is a strong training-free baseline and a practical first step before practitioners pay the cost of data generation, pseudo-label curation, and fine-tuning.

\section{Conclusion}
\label{sec:conclusion}

Our analysis demonstrates that unsupervised PCA-based compression is an effective and training-free mechanism for domain adaptation in dense retrieval systems. It is proven to be competitive even against heavyweight pipelines (Section~\ref{App:ida-comparison}). By identifying principal axes from the variance patterns of the target domain queries, PCA achieves consistent performance gains across diverse models and datasets, particularly through query-only compression that outperforms Query+Document Compression (Section~\ref{sec:mainfindings}). Our findings indicate that moderate dimensionality reduction (preserving 50-90\% of components) frequently enhances retrieval performance (Figure~\ref{fig:trade-off}). Notably, these gains persist even when models exhibit varying degrees of pre-existing domain familiarity (Figure~\ref{fig:similarity_improvement}), underscoring PCA's general utility. This paradigm is especially valuable when labeled domain data is scarce, though practitioners should validate retention ratios for their specific use case. Future work could explore integrating PCA with synthetic data generation or nonlinear manifold learning.

\newpage
\section{Limitations}
\label{sec:limitations}

While our approach demonstrates promising results, several limitations merit consideration. First, the method assumes pre-trained encoders already capture sufficient domain-relevant information--for highly specialized domains with unique terminology, PCA's ability to enhance performance is constrained by the original model's knowledge boundaries. Second, though unsupervised, our technique still requires an adequate amount of unlabeled target domain samples for reliable covariance estimation; insufficient queries may lead to unstable component extraction. 
% Third, the optimal retention ratio proves model- and dataset-dependent, necessitating empirical validation rather than offering universal prescriptive guidelines. We leave the criteria for choosing the optimal retention ratio to future works. Finally, as a linear transformation, PCA may inadequately capture nonlinear semantic relationships in embedding spaces, suggesting potential for exploring nonlinear dimensionality reduction methods in future work.
Third, the optimal retention ratio proves model- and dataset-dependent, necessitating empirical validation rather than offering universal prescriptive guidelines. In practice, we use $r=0.9$ as a conservative default and recommend sweeping a small set of retention ratios when validation data is available, falling back to the original embeddings when
compression hurts performance. Developing automatic criteria for choosing the optimal retention ratio remains an important direction for future work.

\section*{Acknowledgments}
This work is supported by the Office of Naval Research (N00014-24-1-2089). We thank the JHU CLSP community for productive discussions.

\bibliography{anthology,custom}
\bibliographystyle{acl_natbib}

\clearpage

\appendix

\section{General Trend for Eigenvalue distribution}
\label{App:decay}
The curves in Fig.\ref{fig:components} are approximately linear on log--log axes, consistent with a power‑law-like decay.  To move beyond visual evidence, we quantify this trend across all evaluated datasets with the Sentence-T5 embeddings through the Kolmogorov-Smirnov (KS) test~\cite{clauset2009power}.

\paragraph{Procedure:}
For each dataset, we form the sample covariance of mean‑centered embeddings, \(C=\frac{1}{n-1}X_c^\top X_c\), and obtain its eigenvalues \(\lambda_1 \ge \lambda_2 \ge \cdots\). We then fit the tail of the spectrum to a rank--size power law
\[
\lambda_k \approx C\,k^{-\beta}, \qquad k \ge k_{\min},
\]
choosing \(k_{\min}\) automatically to minimize the Kolmogorov-Smirnov (KS) distance between the empirical tail and its fitted curve, with a minimum tail length of 10 (i.e., the default lower bound~\cite{clauset2009power}). The decay exponent \(\beta\) is estimated by ordinary least squares (OLS) on log--log scales,
\[
\log \lambda_k = a - \beta \log k + \varepsilon_k, \quad k \ge k_{\min},
\]
and we report (i) the estimated \(\beta\) with an OLS‑based 95\% confidence interval, (ii) the log--log \(R^2\), and (iii) a KS goodness‑of‑fit \(p\)‑value obtained by parametric bootstrap under the fitted power law.

\paragraph{Summary statistics:}
Across all evaluated spectra, \(\bar{\beta}=82.716 \pm 7.909\), \(R^2=0.971 \pm 0.012\), and every dataset passes the KS check with \(p \ge 0.10\) (100\%). These results indicate a consistently \emph{steep} power‑law--like decay of eigenvalues across datasets, supporting the practical use of aggressive low‑rank compression: a relatively small number of leading components accounts for the bulk of the variance, while later components contribute negligibly.

\paragraph{Remarks:}
This analysis is descriptive rather than generative: eigenvalues are ordered and not independent, and the precise value of \(\hat{\beta}\) can vary with the tail threshold and normalization choice. Nevertheless, the agreement between visual log--log linearity, and the uniformly good KS diagnostics makes the power‑law interpretation a robust summary of the spectra observed; the two datasets in Fig.~\ref{fig:components} serve as representative exemplars.

% \section{Example Appendix}
% \label{sec:appendix}

\section{Generalizability of Domain Semantics}
\label{app:cross-validation}

To verify that our method captures generalizable domain semantics rather than merely overfitting to a specific set of test queries, we conducted a 3-fold cross-validation experiment.

For each dataset, we randomly split the queries into 3 folds. In each run, we fit the PCA only on the observed folds (simulating historical data) and evaluated on the held-out fold (unseen future queries). Table~\ref{tab:cross-validation-results} presents the percentage improvement in NDCG@10 for Query Compression under this strict cross-validation setting.

The results are highly consistent with the results reported in the main paper. For example, NarrativeQA with GTE retains a $\sim$41\% improvement even on unseen queries. This confirms that the principal components learned from a sample of queries successfully encode the underlying \textit{domain structure} (user intent variance) rather than just memorizing the test set.

\begin{table}[h]
\centering
\scriptsize
\renewcommand{\arraystretch}{1.1}
\setlength{\tabcolsep}{1pt}
\resizebox{\columnwidth}{!}{%
\begin{tabular}{l ccccccccc}
\toprule
\textbf{Dataset} & \textbf{MiniLM} & \textbf{GTE} & \textbf{Instr.} & \textbf{Dis-Ro} & \textbf{MP-All} & \textbf{MP-QA} & \textbf{Sent-T5} & \textbf{BGE} & \textbf{SFR} \\
\midrule
\textbf{Apps} & \cellcolor{red!25}-2.6 & \cellcolor{green!25}0.5 & \cellcolor{red!25}-13.8 & \cellcolor{red!25}-13.3 & \cellcolor{green!25}1.3 & \cellcolor{red!25}-11.4 & \cellcolor{green!25}19.9 & \cellcolor{red!25}-14.9 & \cellcolor{red!25}-0.2 \\
\textbf{SciDocs} & \cellcolor{red!25}-1.3 & \cellcolor{green!25}19.3 & \cellcolor{green!25}2.8 & \cellcolor{red!25}-0.3 & \cellcolor{red!25}-1.1 & \cellcolor{green!25}2.9 & \cellcolor{green!25}3.3 & \cellcolor{green!25}4.3 & \cellcolor{green!25}2.0 \\
\textbf{MedQA} & \cellcolor{green!25}2.0 & \cellcolor{green!25}1.8 & \cellcolor{green!25}1.6 & \cellcolor{green!25}2.5 & \cellcolor{green!25}1.7 & \cellcolor{green!25}2.2 & \cellcolor{green!25}6.9 & \cellcolor{green!25}0.8 & \cellcolor{green!25}0.5 \\
\textbf{ArguAna} & \cellcolor{green!25}1.4 & \cellcolor{red!25}-1.7 & \cellcolor{green!25}2.5 & \cellcolor{green!25}0.3 & \cellcolor{red!25}-0.1 & \cellcolor{red!25}-0.2 & \cellcolor{green!25}10.9 & \cellcolor{green!25}2.4 & \cellcolor{red!25}-1.3 \\
\textbf{StackOverflow} & \cellcolor{green!25}0.1 & \cellcolor{red!25}-0.1 & \cellcolor{red!25}-1.1 & \cellcolor{green!25}0.4 & \cellcolor{red!25}-0.1 & \cellcolor{green!25}1.3 & \cellcolor{green!25}2.9 & \cellcolor{green!25}0.6 & \cellcolor{red!25}-0.9 \\
\textbf{TV2Nord} & \cellcolor{green!25}37.5 & \cellcolor{green!25}4.2 & \cellcolor{green!25}2.7 & \cellcolor{green!25}16.1 & \cellcolor{green!25}15.9 & \cellcolor{green!25}7.4 & \cellcolor{green!25}1.5 & \cellcolor{green!25}0.1 & \cellcolor{green!25}5.4 \\
\textbf{MLQA} & \cellcolor{green!25}0.4 & \cellcolor{green!25}0.6 & \cellcolor{green!25}2.2 & \cellcolor{green!25}0.7 & \cellcolor{green!25}1.0 & \cellcolor{green!25}0.6 & \cellcolor{green!25}0.8 & \cellcolor{green!25}0.2 & \cellcolor{green!25}1.8 \\
\textbf{NarrativeQA} & \cellcolor{green!25}10.9 & \cellcolor{green!25}41.8 & \cellcolor{green!25}5.6 & \cellcolor{green!25}16.4 & \cellcolor{green!25}13.5 & \cellcolor{green!25}5.8 & \cellcolor{green!25}20.9 & \cellcolor{green!25}2.2 & \cellcolor{green!25}3.0 \\
\textbf{SpartQA} & \cellcolor{green!25}537.4 & \cellcolor{green!25}124.2 & \cellcolor{green!25}304.7 & \cellcolor{green!25}456.9 & \cellcolor{green!25}1813.6 & \cellcolor{green!25}434.7 & \cellcolor{green!25}110.7 & \cellcolor{green!25}163.6 & \cellcolor{green!25}841.7 \\
\bottomrule
\end{tabular}%
}
\caption{NDCG@10 improvement (\%) for Query Compression using 3-fold cross-validation. Positive gains are highlighted \colorbox{green!25}{green}, negative values are \colorbox{red!25}{red}. This setting evaluates the method on unseen queries to ensure generalization.}
\label{tab:cross-validation-results}
\end{table}

\section{Recall and Precision}
\label{App:rec_and_pre}

Tables~\ref{tab:query-recall-results}, \ref{tab:query-doc-recall-results} and Tables~\ref{tab:query-precision-results}, \ref{tab:query-doc-precision-results} show the results for Recall@10 and Precision@10 similar to Table~\ref{tab:merged-compression-results}. Their performance gain patterns are almost identical with Table~\ref{tab:merged-compression-results}.

\begin{table}[h]
\centering

\resizebox{\columnwidth}{!}{%
\scriptsize
\renewcommand{\arraystretch}{1.1}
\setlength{\tabcolsep}{1pt}
\begin{tabular}{l ccccccccc}
\toprule
\multirow{2}{*}{\textbf{Datasets}} & \multicolumn{9}{c}{\textbf{Query Compression}} \\
\cmidrule(lr){2-10}
& \textbf{MiniLM} & \textbf{GTE} & \textbf{Instr.} & \textbf{Dis-Ro} & \textbf{MP-QA} & \textbf{MP-All} & \textbf{Sent-T5} & \textbf{BGE} & \textbf{SFR} \\
\midrule
\textbf{Code} & \gcell{1.0}{1.0} & \gcell{0.8}{0.8} & \gcell{-0.7}{-0.7} & \gcell{5.0}{5.0} & \gcell{0.97}{0.97} & \gcell{0.85}{0.85} & \gcell{30.4}{30.4} & \gcell{2.5}{2.5} & \gcell{-0.1}{-0.1} \\
\textbf{Apps} & \gcell{-1.6}{-1.6} & \gcell{0.2}{0.2} & \gcell{-12.3}{-12.3} & \gcell{-13.6}{-13.6} & \gcell{-11.4}{-11.4} & \gcell{2.75}{2.75} & \gcell{13.7}{13.7} & \gcell{-13.9}{-13.9} & \gcell{-0.3}{-0.3} \\
\textbf{SciDocs} & \gcell{-0.3}{-0.3} & \gcell{19.2}{19.2} & \gcell{4.4}{4.4} & \gcell{-0.4}{-0.4} & \gcell{2.7}{2.7} & \gcell{-1.2}{-1.2} & \gcell{4.5}{4.5} & \gcell{3.0}{3.0} & \gcell{0.8}{0.8} \\
\textbf{MedQA} & \gcell{1.7}{1.7} & \gcell{1.2}{1.2} & \gcell{1.0}{1.0} & \gcell{1.7}{1.7} & \gcell{2.0}{2.0} & \gcell{1.4}{1.4} & \gcell{5.7}{5.7} & \gcell{1.6}{1.6} & \gcell{0.5}{0.5} \\
\textbf{ArguAna} & \gcell{0.4}{0.4} & \gcell{-1.1}{-1.1} & \gcell{1.3}{1.3} & \gcell{0.6}{0.6} & \gcell{-0.2}{-0.2} & \gcell{-0.2}{-0.2} & \gcell{9.6}{9.6} & \gcell{2.9}{2.9} & \gcell{-0.6}{-0.6} \\
\textbf{StackOverflow} & \gcell{0.1}{0.1} & \gcell{-0.3}{-0.3} & \gcell{-1.0}{-1.0} & \gcell{0.2}{0.2} & \gcell{1.1}{1.1} & \gcell{-0.1}{-0.1} & \gcell{3.2}{3.2} & \gcell{0.1}{0.1} & \gcell{-1.0}{-1.0} \\
\textbf{TV2Nord} & \gcell{33.4}{33.4} & \gcell{3.3}{3.3} & \gcell{2.5}{2.5} & \gcell{15.9}{15.9} & \gcell{7.4}{7.4} & \gcell{15.1}{15.1} & \gcell{1.6}{1.6} & \gcell{0.1}{0.1} & \gcell{5.6}{5.6} \\
\textbf{GerDa} & \gcell{11.2}{11.2} & \gcell{5.4}{5.4} & \gcell{18.0}{18.0} & \gcell{2.1}{2.1} & \gcell{7.1}{7.1} & \gcell{8.0}{8.0} & \gcell{3.4}{3.4} & \gcell{3.0}{3.0} & \gcell{8.4}{8.4} \\
\textbf{ARC} & \gcell{2.5}{2.5} & \gcell{5.0}{5.0} & \gcell{14.3}{14.3} & \gcell{6.5}{6.5} & \gcell{1.8}{1.8} & \gcell{2.9}{2.9} & \gcell{11.6}{11.6} & \gcell{8.2}{8.2} & \gcell{6.5}{6.5} \\
\textbf{FeedbackQA} & \gcell{-0.9}{-0.9} & \gcell{0.7}{0.7} & \gcell{-3.3}{-3.3} & \gcell{0.9}{0.9} & \gcell{-2.7}{-2.7} & \gcell{-1.3}{-1.3} & \gcell{-1.5}{-1.5} & \gcell{-0.3}{-0.3} & \gcell{-0.2}{-0.2} \\
\textbf{FaithDial} & \gcell{0.9}{0.9} & \gcell{4.8}{4.8} & \gcell{5.2}{5.2} & \gcell{-0.1}{-0.1} & \gcell{2.7}{2.7} & \gcell{1.7}{1.7} & \gcell{5.1}{5.1} & \gcell{8.9}{8.9} & \gcell{3.3}{3.3} \\
\textbf{MLQA} & \gcell{0.5}{0.5} & \gcell{0.8}{0.8} & \gcell{2.1}{2.1} & \gcell{0.7}{0.7} & \gcell{0.6}{0.6} & \gcell{1.0}{1.0} & \gcell{-0.2}{-0.2} & \gcell{0.2}{0.2} & \gcell{1.5}{1.5} \\
\textbf{NarrativeQA} & \gcell{10.4}{10.4} & \gcell{39.8}{39.8} & \gcell{3.2}{3.2} & \gcell{18.9}{18.9} & \gcell{4.9}{4.9} & \gcell{15.2}{15.2} & \gcell{20.1}{20.1} & \gcell{2.2}{2.2} & \gcell{3.7}{3.7} \\
\textbf{SpartQA} & \gcell{362.2}{362.2} & \gcell{73.9}{73.9} & \gcell{232.4}{232.4} & \gcell{313.8}{313.8} & \gcell{265.2}{265.2} & \gcell{1207.1}{1207.1} & \gcell{106.8}{106.8} & \gcell{116.5}{116.5} & \gcell{696.3}{696.3} \\
\midrule
\textbf{Summary} & 11/14 & 12/14 & 10/14 & 11/14 & 11/14 & 10/14 & 12/14 & 12/14 & 9/14 \\
\bottomrule
\end{tabular}%
}

\caption{Recall@10 improvement (\%) for Query Compression. Positive gains are highlighted \colorbox{green!25}{green}, negative values are \colorbox{red!25}{red}. This method enhances performance in 98 out of 126 data–model pairs (77.8\%).}
\label{tab:query-recall-results}
\end{table}

\begin{table}[htbp]
\centering

\resizebox{\columnwidth}{!}{%
\scriptsize
\renewcommand{\arraystretch}{1.1}
\setlength{\tabcolsep}{1pt}
\begin{tabular}{l ccccccccc}
\toprule
\multirow{2}{*}{\textbf{Datasets}} & \multicolumn{9}{c}{\textbf{Query+Document Compression}} \\
\cmidrule(lr){2-10}
& \textbf{MiniLM} & \textbf{GTE} & \textbf{Instr.} & \textbf{Dis-Ro} & \textbf{MP-QA} & \textbf{MP-All} & \textbf{Sent-T5} & \textbf{BGE} & \textbf{SFR} \\
\midrule
\textbf{Code} & \bcell{-0.7}{-0.7} & \bcell{-0.1}{-0.1} & \bcell{-1.2}{-1.2} & \bcell{0.5}{0.5} & \bcell{-1.3}{-1.3} & \bcell{-1.0}{-1.0} & \bcell{16.1}{16.1} & \bcell{-1.0}{-1.0} & \bcell{-0.3}{-0.3} \\
\textbf{Apps} & \bcell{-17.5}{-17.5} & \bcell{-1.3}{-1.3} & \bcell{-22.8}{-22.8} & \bcell{-48.1}{-48.1} & \bcell{-23.9}{-23.9} & \bcell{-11.0}{-11.0} & \bcell{4.5}{4.5} & \bcell{-35.0}{-35.0} & \bcell{-1.3}{-1.3} \\
\textbf{SciDocs} & \bcell{-0.0}{-0.0} & \bcell{14.3}{14.3} & \bcell{3.6}{3.6} & \bcell{1.0}{1.0} & \bcell{2.3}{2.3} & \bcell{0.0}{0.0} & \bcell{-4.6}{-4.6} & \bcell{-7.8}{-7.8} & \bcell{0.4}{0.4} \\
\textbf{MedQA} & \bcell{1.6}{1.6} & \bcell{1.0}{1.0} & \bcell{1.3}{1.3} & \bcell{1.6}{1.6} & \bcell{1.6}{1.6} & \bcell{0.9}{0.9} & \bcell{4.6}{4.6} & \bcell{1.4}{1.4} & \bcell{0.6}{0.6} \\
\textbf{ArguAna} & \bcell{0.1}{0.1} & \bcell{-1.3}{-1.3} & \bcell{0.8}{0.8} & \bcell{-0.3}{-0.3} & \bcell{-1.3}{-1.3} & \bcell{-1.5}{-1.5} & \bcell{6.3}{6.3} & \bcell{1.6}{1.6} & \bcell{-0.9}{-0.9} \\
\textbf{StackOverflow} & \bcell{0.0}{0.0} & \bcell{-0.3}{-0.3} & \bcell{-1.1}{-1.1} & \bcell{-0.7}{-0.7} & \bcell{0.3}{0.3} & \bcell{-0.1}{-0.1} & \bcell{3.2}{3.2} & \bcell{-0.3}{-0.3} & \bcell{-1.0}{-1.0} \\
\textbf{TV2Nord} & \bcell{20.7}{20.7} & \bcell{2.3}{2.3} & \bcell{1.8}{1.8} & \bcell{9.3}{9.3} & \bcell{3.4}{3.4} & \bcell{3.4}{3.4} & \bcell{0.0}{0.0} & \bcell{0.0}{0.0} & \bcell{2.8}{2.8} \\
\textbf{GerDa} & \bcell{1.1}{1.1} & \bcell{-9.3}{-9.3} & \bcell{3.9}{3.9} & \bcell{-8.4}{-8.4} & \bcell{-1.2}{-1.2} & \bcell{-2.1}{-2.1} & \bcell{-9.4}{-9.4} & \bcell{-11.9}{-11.9} & \bcell{-11.8}{-11.8} \\
\textbf{ARC} & \bcell{5.0}{5.0} & \bcell{5.8}{5.8} & \bcell{12.8}{12.8} & \bcell{5.1}{5.1} & \bcell{2.3}{2.3} & \bcell{2.5}{2.5} & \bcell{8.5}{8.5} & \bcell{7.6}{7.6} & \bcell{3.3}{3.3} \\
\textbf{FeedbackQA} & \bcell{-1.9}{-1.9} & \bcell{0.8}{0.8} & \bcell{-5.1}{-5.1} & \bcell{-6.7}{-6.7} & \bcell{-4.6}{-4.6} & \bcell{-3.1}{-3.1} & \bcell{-3.5}{-3.5} & \bcell{-0.7}{-0.7} & \bcell{-0.7}{-0.7} \\
\textbf{FaithDial} & \bcell{1.6}{1.6} & \bcell{4.7}{4.7} & \bcell{3.6}{3.6} & \bcell{-0.1}{-0.1} & \bcell{1.6}{1.6} & \bcell{1.8}{1.8} & \bcell{4.1}{4.1} & \bcell{8.2}{8.2} & \bcell{3.4}{3.4} \\
\textbf{MLQA} & \bcell{0.4}{0.4} & \bcell{0.8}{0.8} & \bcell{2.1}{2.1} & \bcell{0.3}{0.3} & \bcell{0.4}{0.4} & \bcell{0.9}{0.9} & \bcell{-1.3}{-1.3} & \bcell{0.4}{0.4} & \bcell{1.2}{1.2} \\
\textbf{NarrativeQA} & \bcell{7.0}{7.0} & \bcell{35.9}{35.9} & \bcell{0.5}{0.5} & \bcell{14.5}{14.5} & \bcell{-0.1}{-0.1} & \bcell{10.8}{10.8} & \bcell{17.0}{17.0} & \bcell{2.0}{2.0} & \bcell{2.9}{2.9} \\
\textbf{SpartQA} & \bcell{371.1}{371.1} & \bcell{86.2}{86.2} & \bcell{492.9}{492.9} & \bcell{281.6}{281.6} & \bcell{340.0}{340.0} & \bcell{1035.7}{1035.7} & \bcell{214.9}{214.9} & \bcell{118.1}{118.1} & \bcell{713.8}{713.8} \\
\midrule
\textbf{Summary} & 9/14 & 9/14 & 10/14 & 8/14 & 8/14 & 7/14 & 9/14 & 7/14 & 8/14 \\
\bottomrule
\end{tabular}%
}
\caption{Recall@10 improvement (\%) for Query+Document Compression. Positive gains are highlighted \colorbox{blue!25}{blue}, negative values are \colorbox{red!25}{red}. This method enhances performance in 75 out of 126 data–model pairs (59.5\%).}
\label{tab:query-doc-recall-results}

\end{table}

\begin{table}[h]
\centering

\resizebox{\columnwidth}{!}{%
\scriptsize
\renewcommand{\arraystretch}{1.1}
\setlength{\tabcolsep}{1pt}
\begin{tabular}{l ccccccccc}
\toprule
\multirow{2}{*}{\textbf{Datasets}} & \multicolumn{9}{c}{\textbf{Query Compression}} \\
\cmidrule(lr){2-10}
& \textbf{MiniLM} & \textbf{GTE} & \textbf{Instr.} & \textbf{Dis-Ro} & \textbf{MP-QA} & \textbf{MP-All} & \textbf{Sent-T5} & \textbf{BGE} & \textbf{SFR} \\
\midrule
\textbf{Code} & \gcell{0.9}{0.9} & \gcell{0.8}{0.8} & \gcell{-0.7}{-0.7} & \gcell{5.0}{5.0} & \gcell{0.9}{0.9} & \gcell{0.8}{0.8} & \gcell{30.4}{30.4} & \gcell{2.4}{2.4} & \gcell{-0.1}{-0.1} \\
\textbf{Apps} & \gcell{-2.0}{-2.0} & \gcell{0.2}{0.2} & \gcell{-12.3}{-12.3} & \gcell{-14.0}{-14.0} & \gcell{-11.2}{-11.2} & \gcell{2.4}{2.4} & \gcell{14.1}{14.1} & \gcell{-13.7}{-13.7} & \gcell{-0.3}{-0.3} \\
\textbf{SciDocs} & \gcell{-0.2}{-0.2} & \gcell{18.9}{18.9} & \gcell{4.3}{4.3} & \gcell{-0.4}{-0.4} & \gcell{2.8}{2.8} & \gcell{-1.1}{-1.1} & \gcell{4.3}{4.3} & \gcell{2.8}{2.8} & \gcell{0.8}{0.8} \\
\textbf{MedQA} & \gcell{1.6}{1.6} & \gcell{1.2}{1.2} & \gcell{1.0}{1.0} & \gcell{1.7}{1.7} & \gcell{1.9}{1.9} & \gcell{1.4}{1.4} & \gcell{5.7}{5.7} & \gcell{1.5}{1.5} & \gcell{0.5}{0.5} \\
\textbf{ArguAna} & \gcell{0.5}{0.5} & \gcell{-1.0}{-1.0} & \gcell{1.2}{1.2} & \gcell{0.6}{0.6} & \gcell{-0.1}{-0.1} & \gcell{-0.1}{-0.1} & \gcell{9.4}{9.4} & \gcell{2.9}{2.9} & \gcell{-0.6}{-0.6} \\
\textbf{StackOverflow} & \gcell{0.1}{0.1} & \gcell{-0.3}{-0.3} & \gcell{-0.9}{-0.9} & \gcell{0.2}{0.2} & \gcell{0.9}{0.9} & \gcell{-0.1}{-0.1} & \gcell{3.1}{3.1} & \gcell{0.1}{0.1} & \gcell{-0.9}{-0.9} \\
\textbf{TV2Nord} & \gcell{33.3}{33.3} & \gcell{3.3}{3.3} & \gcell{2.3}{2.3} & \gcell{15.7}{15.7} & \gcell{7.5}{7.5} & \gcell{14.9}{14.9} & \gcell{1.6}{1.6} & \gcell{0.0}{0.0} & \gcell{5.5}{5.5} \\
\textbf{GerDa} & \gcell{12.5}{12.5} & \gcell{5.6}{5.6} & \gcell{17.8}{17.8} & \gcell{2.7}{2.7} & \gcell{6.9}{6.9} & \gcell{7.5}{7.5} & \gcell{3.2}{3.2} & \gcell{2.9}{2.9} & \gcell{7.7}{7.7} \\
\textbf{ARC} & \gcell{2.3}{2.3} & \gcell{5.0}{5.0} & \gcell{14.4}{14.4} & \gcell{6.5}{6.5} & \gcell{2.1}{2.1} & \gcell{2.9}{2.9} & \gcell{11.4}{11.4} & \gcell{8.2}{8.2} & \gcell{6.7}{6.7} \\
\textbf{FeedbackQA} & \gcell{-0.9}{-0.9} & \gcell{0.7}{0.7} & \gcell{-3.3}{-3.3} & \gcell{0.8}{0.8} & \gcell{-2.6}{-2.6} & \gcell{-1.2}{-1.2} & \gcell{-1.5}{-1.5} & \gcell{-0.2}{-0.2} & \gcell{-0.2}{-0.2} \\
\textbf{FaithDial} & \gcell{0.8}{0.8} & \gcell{4.7}{4.7} & \gcell{5.3}{5.3} & \gcell{-0.2}{-0.2} & \gcell{2.6}{2.6} & \gcell{1.5}{1.5} & \gcell{5.3}{5.3} & \gcell{8.6}{8.6} & \gcell{3.2}{3.2} \\
\textbf{MLQA} & \gcell{0.5}{0.5} & \gcell{0.7}{0.7} & \gcell{2.0}{2.0} & \gcell{0.6}{0.6} & \gcell{0.5}{0.5} & \gcell{1.0}{1.0} & \gcell{-0.2}{-0.2} & \gcell{0.2}{0.2} & \gcell{1.5}{1.5} \\
\textbf{NarrativeQA} & \gcell{10.4}{10.4} & \gcell{39.8}{39.8} & \gcell{3.4}{3.4} & \gcell{18.7}{18.7} & \gcell{5.0}{5.0} & \gcell{15.1}{15.1} & \gcell{20.2}{20.2} & \gcell{2.2}{2.2} & \gcell{3.5}{3.5} \\
\textbf{SpartQA} & \gcell{428.3}{428.3} & \gcell{97.9}{97.9} & \gcell{329.1}{329.1} & \gcell{390.1}{390.1} & \gcell{300.0}{300.0} & \gcell{1433.3}{1433.3} & \gcell{84.0}{84.0} & \gcell{148.0}{148.0} & \gcell{784.6}{784.6} \\
\midrule
\textbf{Summary} & 11/14 & 12/14 & 10/14 & 11/14 & 11/14 & 10/14 & 12/14 & 11/14 & 9/14 \\
\bottomrule
\end{tabular}%
}
\caption{Precision@10 improvement (\%) for Query Compression. Positive gains are highlighted \colorbox{green!25}{green}, negative values are \colorbox{red!25}{red}. This method enhances performance in 97 out of 126 data–model pairs (77.0\%).}
\label{tab:query-precision-results}
\end{table}

\begin{table}[h]
\centering

\resizebox{\columnwidth}{!}{%
\scriptsize
\renewcommand{\arraystretch}{1.1}
\setlength{\tabcolsep}{1pt}
\begin{tabular}{l ccccccccc}
\toprule
\multirow{2}{*}{\textbf{Datasets}} & \multicolumn{9}{c}{\textbf{Query+Document Compression}} \\
\cmidrule(lr){2-10}
& \textbf{MiniLM} & \textbf{GTE} & \textbf{Instr.} & \textbf{Dis-Ro} & \textbf{MP-QA} & \textbf{MP-All} & \textbf{Sent-T5} & \textbf{BGE} & \textbf{SFR} \\
\midrule
\textbf{Code} & \bcell{-0.7}{-0.7} & \bcell{0.0}{0.0} & \bcell{-1.1}{-1.1} & \bcell{0.5}{0.5} & \bcell{-1.2}{-1.2} & \bcell{-1.0}{-1.0} & \bcell{16.1}{16.1} & \bcell{-1.0}{-1.0} & \bcell{-0.3}{-0.3} \\
\textbf{Apps} & \bcell{-17.1}{-17.1} & \bcell{-1.2}{-1.2} & \bcell{-22.4}{-22.4} & \bcell{-48.4}{-48.4} & \bcell{-24.1}{-24.1} & \bcell{-11.2}{-11.2} & \bcell{5.0}{5.0} & \bcell{-35.0}{-35.0} & \bcell{-1.2}{-1.2} \\
\textbf{SciDocs} & \bcell{0.0}{0.0} & \bcell{14.1}{14.1} & \bcell{3.6}{3.6} & \bcell{0.9}{0.9} & \bcell{2.4}{2.4} & \bcell{0.0}{0.0} & \bcell{-4.6}{-4.6} & \bcell{-7.7}{-7.7} & \bcell{0.3}{0.3} \\
\textbf{MedQA} & \bcell{1.6}{1.6} & \bcell{1.0}{1.0} & \bcell{1.3}{1.3} & \bcell{1.6}{1.6} & \bcell{1.5}{1.5} & \bcell{0.8}{0.8} & \bcell{4.5}{4.5} & \bcell{1.3}{1.3} & \bcell{0.5}{0.5} \\
\textbf{ArguAna} & \bcell{0.1}{0.1} & \bcell{-1.2}{-1.2} & \bcell{0.7}{0.7} & \bcell{-0.2}{-0.2} & \bcell{-1.2}{-1.2} & \bcell{-1.4}{-1.4} & \bcell{6.2}{6.2} & \bcell{1.7}{1.7} & \bcell{-0.9}{-0.9} \\
\textbf{StackOverflow} & \bcell{0.0}{0.0} & \bcell{-0.3}{-0.3} & \bcell{-1.0}{-1.0} & \bcell{-0.6}{-0.6} & \bcell{0.2}{0.2} & \bcell{-0.1}{-0.1} & \bcell{3.1}{3.1} & \bcell{-0.2}{-0.2} & \bcell{-1.0}{-1.0} \\
\textbf{TV2Nord} & \bcell{20.7}{20.7} & \bcell{2.3}{2.3} & \bcell{1.7}{1.7} & \bcell{9.2}{9.2} & \bcell{3.5}{3.5} & \bcell{3.2}{3.2} & \bcell{0.0}{0.0} & \bcell{0.0}{0.0} & \bcell{2.7}{2.7} \\
\textbf{GerDa} & \bcell{2.5}{2.5} & \bcell{-9.3}{-9.3} & \bcell{3.6}{3.6} & \bcell{-6.9}{-6.9} & \bcell{-2.3}{-2.3} & \bcell{-1.5}{-1.5} & \bcell{-9.0}{-9.0} & \bcell{-11.7}{-11.7} & \bcell{-12.2}{-12.2} \\
\textbf{ARC} & \bcell{4.7}{4.7} & \bcell{5.9}{5.9} & \bcell{12.6}{12.6} & \bcell{5.4}{5.4} & \bcell{2.6}{2.6} & \bcell{2.4}{2.4} & \bcell{8.5}{8.5} & \bcell{7.5}{7.5} & \bcell{3.3}{3.3} \\
\textbf{FeedbackQA} & \bcell{-1.9}{-1.9} & \bcell{0.8}{0.8} & \bcell{-5.1}{-5.1} & \bcell{-6.7}{-6.7} & \bcell{-4.6}{-4.6} & \bcell{-3.0}{-3.0} & \bcell{-3.5}{-3.5} & \bcell{-0.7}{-0.7} & \bcell{-0.7}{-0.7} \\
\textbf{FaithDial} & \bcell{1.7}{1.7} & \bcell{4.5}{4.5} & \bcell{3.5}{3.5} & \bcell{-0.2}{-0.2} & \bcell{1.5}{1.5} & \bcell{1.7}{1.7} & \bcell{4.2}{4.2} & \bcell{7.9}{7.9} & \bcell{3.4}{3.4} \\
\textbf{MLQA} & \bcell{0.5}{0.5} & \bcell{0.7}{0.7} & \bcell{2.0}{2.0} & \bcell{0.2}{0.2} & \bcell{0.2}{0.2} & \bcell{0.9}{0.9} & \bcell{-1.3}{-1.3} & \bcell{0.3}{0.3} & \bcell{1.2}{1.2} \\
\textbf{NarrativeQA} & \bcell{7.2}{7.2} & \bcell{35.8}{35.8} & \bcell{0.5}{0.5} & \bcell{14.1}{14.1} & \bcell{0.0}{0.0} & \bcell{10.6}{10.6} & \bcell{17.0}{17.0} & \bcell{2.0}{2.0} & \bcell{2.9}{2.9} \\
\textbf{SpartQA} & \bcell{456.6}{456.6} & \bcell{115.0}{115.0} & \bcell{697.9}{697.9} & \bcell{373.7}{373.7} & \bcell{373.5}{373.5} & \bcell{1255.5}{1255.5} & \bcell{195.6}{195.6} & \bcell{151.5}{151.5} & \bcell{810.2}{810.2} \\
\midrule
\textbf{Summary} & 9/14 & 9/14 & 10/14 & 8/14 & 8/14 & 7/14 & 9/14 & 7/14 & 8/14 \\
\bottomrule
\end{tabular}%
}
\caption{Precision@10 improvement (\%) for Query+Document Compression. Positive gains are highlighted \colorbox{blue!25}{blue}, negative values are \colorbox{red!25}{red}. This method enhances performance in 75 out of 126 data–model pairs (59.5\%).}
\label{tab:query-doc-precision-results}
\end{table}

\section{Random Compression Baseline}

To verify that the improvements in Table~\ref{tab:merged-compression-results} are not by chance, we also experiment on 90\% Random Compression (Table~\ref{tab:random-compression-results}), where 10\% of the original embedding dimensions are randomly chosen and removed. It confirms that Random Compression is unlikely to improve retrieval performance, and the gains from Query Compression are not by chance.

\begin{table}[htbp]
\centering
\scriptsize
\renewcommand{\arraystretch}{1.1}
\setlength{\tabcolsep}{1pt}
\resizebox{\columnwidth}{!}{%
\begin{tabular}{l ccccccccc}
\toprule
\multirow{2}{*}{\textbf{Datasets}} & \multicolumn{9}{c}{\textbf{90\% Random Compression}} \\
\cmidrule(lr){2-10}
 & \textbf{MiniLM} & \textbf{GTE} & \textbf{Instr.} & \textbf{Dis-Ro} & \textbf{MP-All} & \textbf{MP-QA} & \textbf{Sent-T5} & \textbf{BGE} & \textbf{SFR} \\
\midrule
\textbf{Code}        & \rcell{-0.23}{-0.23} & \rcell{-0.18}{-0.18} & \rcell{-0.03}{-0.03} & \rcell{-0.12}{-0.12} & \rcell{-0.13}{-0.13} & \rcell{-0.02}{-0.02} & \rcell{-0.94}{-0.94} & \gcell{0.07}{0.07} & \rcell{-0.18}{-0.18} \\
\textbf{Apps}        & \rcell{-2.58}{-2.58} & \rcell{-0.23}{-0.23} & \rcell{-1.87}{-1.87} & \gcell{0.81}{0.81} & \rcell{-0.83}{-0.83} & \rcell{-0.78}{-0.78} & \gcell{0.76}{0.76} & \rcell{-2.17}{-2.17} & \rcell{-0.67}{-0.67} \\
\textbf{SciDocs}     & \rcell{-1.2}{-1.2} & \rcell{-1.95}{-1.95} & \gcell{0.22}{0.22} & \rcell{-0.09}{-0.09} & \rcell{-0.76}{-0.76} & \rcell{-0.75}{-0.75} & \rcell{-0.31}{-0.31} & \gcell{0.37}{0.37} & \gcell{1.74}{1.74} \\
\textbf{MedQA}       & \gcell{0.65}{0.65} & \rcell{-0.31}{-0.31} & \gcell{0.18}{0.18} & \rcell{-0.22}{-0.22} & \gcell{0.68}{0.68} & \gcell{0.02}{0.02} & \rcell{-0.28}{-0.28} & \rcell{-0.65}{-0.65} & \rcell{-0.46}{-0.46} \\
\textbf{ArguAna}     & \rcell{-0.48}{-0.48} & \rcell{-0.21}{-0.21} & \rcell{-1.04}{-1.04} & \rcell{-0.65}{-0.65} & \rcell{-0.8}{-0.8} & \rcell{-0.45}{-0.45} & \rcell{-1.29}{-1.29} & \rcell{-0.43}{-0.43} & \rcell{-0.99}{-0.99} \\
\textbf{StackOverflow} & \rcell{-0.48}{-0.48} & \rcell{-0.15}{-0.15} & \rcell{-0.09}{-0.09} & \rcell{-0.39}{-0.39} & \rcell{-0.13}{-0.13} & \rcell{-0.32}{-0.32} & \rcell{-0.12}{-0.12} & \rcell{-0.15}{-0.15} & \rcell{-0.73}{-0.73} \\
\textbf{TV2Nord}    & \rcell{-3.78}{-3.78} & \rcell{-0.91}{-0.91} & \rcell{-0.52}{-0.52} & \rcell{-0.18}{-0.18} & \rcell{-0.69}{-0.69} & \rcell{-0.38}{-0.38} & \rcell{-0.38}{-0.38} & \rcell{-0.32}{-0.32} & \rcell{-1.21}{-1.21} \\
\textbf{GerDa}       & \rcell{0}{0} & \rcell{-1.5}{-1.5} & \rcell{-0.27}{-0.27} & \rcell{-0.23}{-0.23} & \rcell{-1.06}{-1.06} & \rcell{-0.39}{-0.39} & \rcell{-0.14}{-0.14} & \rcell{-0.5}{-0.5} & \rcell{-6.22}{-6.22} \\
\textbf{ARC}         & \rcell{-0.53}{-0.53} & \rcell{-2.18}{-2.18} & \rcell{-0.72}{-0.72} & \rcell{-0.96}{-0.96} & \gcell{0.85}{0.85} & \gcell{0.09}{0.09} & \rcell{-0.12}{-0.12} & \rcell{-2.22}{-2.22} & \gcell{2.5}{2.5} \\
\textbf{FeedbackQA}  & \gcell{0.03}{0.03} & \rcell{-0.49}{-0.49} & \rcell{-0.11}{-0.11} & \gcell{0.12}{0.12} & \rcell{-0.37}{-0.37} & \rcell{-0.03}{-0.03} & \rcell{-0.73}{-0.73} & \rcell{-0.13}{-0.13} & \rcell{-0.65}{-0.65} \\
\textbf{FaithDial}   & \rcell{-0.25}{-0.25} & \rcell{-0.35}{-0.35} & \rcell{-0.43}{-0.43} & \rcell{-0.12}{-0.12} & \gcell{0.41}{0.41} & \rcell{-0.08}{-0.08} & \gcell{0.34}{0.34} & \gcell{0.25}{0.25} & \gcell{0.88}{0.88} \\
\textbf{MLQA} & \rcell{-0.62}{-0.62} & \rcell{-0.33}{-0.33} & \rcell{-0.54}{-0.54} & \rcell{-0.44}{-0.44} & \rcell{-0.32}{-0.32} & \rcell{-0.26}{-0.26} & \rcell{-0.58}{-0.58} & \rcell{-0.54}{-0.54} & \gcell{1.51}{1.51} \\
\textbf{NarrativeQA} & \rcell{-2.42}{-2.42} & \rcell{-0.96}{-0.96} & \rcell{-0.54}{-0.54} & \gcell{0.19}{0.19} & \gcell{1.04}{1.04} & \gcell{2.68}{2.68} & \rcell{-2.43}{-2.43} & \rcell{-0.64}{-0.64} & \rcell{-0.69}{-0.69} \\
\textbf{SpartQA}     & \gcell{3.03}{3.03} & \rcell{-0.64}{-0.64} & \gcell{14.56}{14.56} & \rcell{-16.76}{-16.76} & \rcell{-18.18}{-18.18} & \rcell{-3.94}{-3.94} & \gcell{8.17}{8.17} & \gcell{8.54}{8.54} & \rcell{-27.12}{-27.12} \\
\midrule
\textbf{Summary} & 3/14 & 0/14 & 3/14 & 3/14 & 4/14 & 3/14 & 3/14 & 4/14 & 4/14 \\
\bottomrule
\end{tabular}%
}
\caption{NDCG@10 performance (\%) using 90\% Random Compression. Positive values are highlighted \colorbox{green!25}{green} and negative values are \colorbox{red!25}{red}. The ``Summary'' row indicates the number of datasets with improvements per model.}
\label{tab:random-compression-results}
\end{table}

\section{Low-Query Regime Analysis}
\label{app:low}

\begin{table}[h]
\centering
\scriptsize
\renewcommand{\arraystretch}{1}
\setlength{\tabcolsep}{1pt}
\resizebox{\columnwidth}{!}{%
\begin{tabular}{l ccccccccc}
\toprule
\multirow{2}{*}{\textbf{Datasets}} & \multicolumn{9}{c}{\textbf{Query Compression}} \\
\cmidrule(lr){2-10}
 & \textbf{MiniLM} & \textbf{GTE} & \textbf{Instr.} & \textbf{Dis-Ro} & \textbf{MP-QA} & \textbf{MP-All} & \textbf{Sent-T5} & \textbf{BGE} & \textbf{SFR} \\
\midrule
\textbf{CosQA}      & \gcell{1.3}{1.3}   & \gcell{-1.6}{-1.6} & \gcell{0.7}{0.7}   & \gcell{10.4}{10.4} & \gcell{4.5}{4.5}   & \gcell{6.3}{6.3}   & \gcell{7.0}{7.0}   & \gcell{7.7}{7.7}   & \gcell{7.0}{7.0}   \\
\textbf{SciFact}    & \gcell{-0.4}{-0.4}  & \gcell{-0.1}{-0.1} & \gcell{1.4}{1.4}   & \gcell{-0.8}{-0.8}  & \gcell{-1.2}{-1.2}  & \gcell{0.2}{0.2}   & \gcell{2.1}{2.1}   & \gcell{1.7}{1.7}   & \gcell{-2.1}{-2.1} \\
\textbf{NFCorpus}   & \gcell{-2.7}{-2.7}  & \gcell{0.3}{0.3}   & \gcell{-0.1}{-0.1}  & \gcell{-2.8}{-2.8}  & \gcell{-4.7}{-4.7}  & \gcell{-1.9}{-1.9}  & \gcell{0.5}{0.5}   & \gcell{-1.1}{-1.1}  & \gcell{-0.2}{-0.2} \\
\textbf{FiQA}       & \gcell{0.0}{0.0}   & \gcell{-0.2}{-0.2} & \gcell{2.5}{2.5}   & \gcell{-1.0}{-1.0}  & \gcell{-0.6}{-0.6}  & \gcell{-0.1}{-0.1}  & \gcell{0.3}{0.3}   & \gcell{-1.0}{-1.0}  & \gcell{-2.2}{-2.2} \\
\textbf{BuiltBench} & \gcell{0.1}{0.1}   & \gcell{1.6}{1.6}   & \gcell{3.6}{3.6}   & \gcell{1.5}{1.5}   & \gcell{2.3}{2.3}   & \gcell{2.7}{2.7}   & \gcell{2.9}{2.9}   & \gcell{4.3}{4.3}   & \gcell{6.3}{6.3}   \\
\textbf{Legal}      & \gcell{4.7}{4.7}   & \gcell{24.4}{24.4} & \gcell{22.3}{22.3} & \gcell{1.8}{1.8}   & \gcell{27.9}{27.9} & \gcell{9.1}{9.1}   & \gcell{8.9}{8.9}   & \gcell{14.6}{14.6} & \gcell{20.0}{20.0} \\
\textbf{AILACDocs}  & \gcell{19.7}{19.7} & \gcell{29.8}{29.8} & \gcell{0.4}{0.4}   & \gcell{20.3}{20.3} & \gcell{28.4}{28.4} & \gcell{14.7}{14.7} & \gcell{11.5}{11.5} & \gcell{8.5}{8.5}   & \gcell{31.5}{31.5} \\
\textbf{CodeTrans}  & \gcell{-6.4}{-6.4}  & \gcell{-10.9}{-10.9}& \gcell{-3.2}{-3.2}  & \gcell{-1.5}{-1.5}  & \gcell{0.7}{0.7}   & \gcell{-0.5}{-0.5}  & \gcell{-8.2}{-8.2}  & \gcell{-0.8}{-0.8}  & \gcell{-5.1}{-5.1} \\
\textbf{AILASt}     & \gcell{40.7}{40.7} & \gcell{19.1}{19.1} & \gcell{53.6}{53.6} & \gcell{111.2}{111.2}& \gcell{58.4}{58.4} & \gcell{70.9}{70.9} & \gcell{51.9}{51.9} & \gcell{32.0}{32.0} & \gcell{63.5}{63.5} \\
\textbf{COVID}      & \gcell{8.9}{8.9}   & \gcell{14.2}{14.2} & \gcell{7.2}{7.2}   & \gcell{-7.5}{-7.5}  & \gcell{0.8}{0.8}   & \gcell{2.8}{2.8}   & \gcell{-22.8}{-22.8}& \gcell{-6.9}{-6.9}  & \gcell{-11.9}{-11.9}\\
\textbf{ChemNQ}     & \gcell{-10.8}{-10.8} & \gcell{-18.7}{-18.7}& \gcell{3.7}{3.7}   & \gcell{-14.8}{-14.8} & \gcell{-19.5}{-19.5} & \gcell{-19.6}{-19.6}& \gcell{-14.5}{-14.5}& \gcell{-24.7}{-24.7}& \gcell{-11.2}{-11.2}\\
\midrule
\textbf{Summary} & 6/11 & 6/11 & 9/11 & 5/11 & 7/11 & 7/11 & 7/11 & 6/11 & 5/11 \\
\bottomrule
\end{tabular}%
}
\caption{NDCG@10 improvement (\%) comparison for Query Compression for datasets where the number of queries is lower than most model's target PCA dimension. Positive gains are highlighted \colorbox{green!25}{green}, negative values are \colorbox{red!25}{red}. Summary rows show the number of datasets with improvements per model.}
\label{tab:query-low-queries-datasets-results}
\end{table}

\begin{table}[h]
\centering
\scriptsize
\renewcommand{\arraystretch}{1}
\setlength{\tabcolsep}{1pt}
\resizebox{\columnwidth}{!}{%
\begin{tabular}{l ccccccccc}
\toprule
\multirow{2}{*}{\textbf{Datasets}} & \multicolumn{9}{c}{\textbf{Query+Document Compression}} \\
\cmidrule(lr){2-10}
 & \textbf{MiniLM} & \textbf{GTE} & \textbf{Instr.} & \textbf{Dis-Ro} & \textbf{MP-QA} & \textbf{MP-All} & \textbf{Sent-T5} & \textbf{BGE} & \textbf{SFR} \\
\midrule
\textbf{CosQA}      & \bcell{2.5}{2.5} & \bcell{-1.5}{-1.5} & \bcell{5.0}{5.0} & \bcell{4.1}{4.1} & \bcell{8.8}{8.8} & \bcell{-0.4}{-0.4} & \bcell{-1.2}{-1.2} & \bcell{-2.9}{-2.9} & \bcell{0.7}{0.7} \\
\textbf{SciFact}    & \bcell{-1.3}{-1.3} & \bcell{0.4}{0.4} & \bcell{0.5}{0.5} & \bcell{-2.9}{-2.9} & \bcell{-2.3}{-2.3} & \bcell{-0.9}{-0.9} & \bcell{-1.1}{-1.1} & \bcell{-0.8}{-0.8} & \bcell{-1.2}{-1.2} \\
\textbf{NFCorpus}   & \bcell{-7.8}{-7.8} & \bcell{0.0}{0.0} & \bcell{-3.9}{-3.9} & \bcell{-9.9}{-9.9} & \bcell{-3.3}{-3.3} & \bcell{-8.5}{-8.5} & \bcell{-4.8}{-4.8} & \bcell{-4.4}{-4.4} & \bcell{-5.1}{-5.1} \\
\textbf{FiQA}       & \bcell{-0.2}{-0.2} & \bcell{0.1}{0.1} & \bcell{2.9}{2.9} & \bcell{-0.7}{-0.7} & \bcell{-0.5}{-0.5} & \bcell{0.1}{0.1} & \bcell{-2.2}{-2.2} & \bcell{-1.6}{-1.6} & \bcell{-1.2}{-1.2} \\
\textbf{BuiltBench} & \bcell{0.0}{0.0} & \bcell{-0.3}{-0.3} & \bcell{0.6}{0.6} & \bcell{0.4}{0.4} & \bcell{-0.4}{-0.4} & \bcell{0.2}{0.2} & \bcell{-1.2}{-1.2} & \bcell{-0.6}{-0.6} & \bcell{1.5}{1.5} \\
\textbf{Legal}      & \bcell{6.2}{6.2} & \bcell{10.1}{10.1} & \bcell{-5.5}{-5.5} & \bcell{-11.1}{-11.1} & \bcell{-4.6}{-4.6} & \bcell{4.3}{4.3} & \bcell{-11.6}{-11.6} & \bcell{7.1}{7.1} & \bcell{-3.5}{-3.5} \\
\textbf{AILACDocs}  & \bcell{12.2}{12.2} & \bcell{26.1}{26.1} & \bcell{4.2}{4.2} & \bcell{16.3}{16.3} & \bcell{12.2}{12.2} & \bcell{25.1}{25.1} & \bcell{8.9}{8.9} & \bcell{-1.3}{-1.3} & \bcell{22.7}{22.7} \\
\textbf{CodeTrans}  & \bcell{-4.8}{-4.8} & \bcell{0.4}{0.4} & \bcell{-2.7}{-2.7} & \bcell{4.4}{4.4} & \bcell{1.6}{1.6} & \bcell{1.6}{1.6} & \bcell{-2.2}{-2.2} & \bcell{1.5}{1.5} & \bcell{-0.2}{-0.2} \\
\textbf{AILASt}     & \bcell{11.9}{11.9} & \bcell{-0.3}{-0.3} & \bcell{23.8}{23.8} & \bcell{38.6}{38.6} & \bcell{38.5}{38.5} & \bcell{29.1}{29.1} & \bcell{5.4}{5.4} & \bcell{7.1}{7.1} & \bcell{32.8}{32.8} \\
\textbf{COVID}      & \bcell{-2.1}{-2.1} & \bcell{16.9}{16.9} & \bcell{7.0}{7.0} & \bcell{-4.2}{-4.2} & \bcell{-2.6}{-2.6} & \bcell{-2.0}{-2.0} & \bcell{-27.2}{-27.2} & \bcell{-20.9}{-20.9} & \bcell{-2.5}{-2.5} \\
\textbf{ChemNQ}     & \bcell{-0.0}{-0.0} & \bcell{10.3}{10.3} & \bcell{2.3}{2.3} & \bcell{-3.6}{-3.6} & \bcell{0.3}{0.3} & \bcell{1.7}{1.7} & \bcell{1.6}{1.6} & \bcell{-5.0}{-5.0} & \bcell{1.2}{1.2} \\
\midrule
\textbf{Summary} & 4/11 & 7/11 & 8/11 & 5/11 & 5/11 & 7/11 & 3/11 & 3/11 & 5/11 \\
\bottomrule
\end{tabular}%
}
\caption{NDCG@10 improvement (\%) comparison for Query+Document Compression for datasets where the number of queries is lower than most models' target PCA dimension. Positive gains are highlighted \colorbox{blue!25}{blue}, negative values are \colorbox{red!25}{red}. Summary rows show the number of datasets with improvements per model.}
\label{tab:query-doc-low-queries-datasets-results}
\end{table}

The datasets in Table~\ref{tab:query-low-queries-datasets-results} and ~\ref{tab:query-doc-low-queries-datasets-results} with limited queries (typically much lower than the original 768-dimensional embeddings, as suggested by the table caption referring to scenarios where query count is less than the target PCA dimension) reveal PCA's adaptive robustness under Query Compression. The legal QA dataset \emph{AILASt} shows strong positive gains across all nine listed models (ranging from +19.1\% for \texttt{GTE} to +111.2\% for \texttt{Dis-Ro}), indicating PCA effectively isolates structured variance even under severe rank constraints. 

For datasets with extremely few queries, the challenge for PCA becomes more pronounced. For instance, the chemistry dataset \emph{ChemNQ}, with only 27 queries, saw most models exhibit performance drops under Query Compression. This suggests that such a sparse dataset may not provide sufficient information for PCA to effectively identify and emphasize task-relevant variance across most models, with \texttt{Instr.} being an exception showing a marginal improvement (+3.7\%). 

Similarly, \emph{COVID}, with only 50 queries, presented a challenging scenario. However, it's noteworthy that despite this limitation, a majority of models (five out of nine, including \texttt{GTE} +14.2\%, \texttt{MiniLM} +8.9\%, and \texttt{Instr.} +7.2\%) still achieved performance gains with Query Compression. This highlights PCA's potential utility even when query data is scarce. Nevertheless, specific models like \texttt{Sent-T5} (-22.8\%) still showed significant performance degradation, indicating that for some architectures or pre-training regimes, compression with very limited data can be detrimental if crucial high-dimensional interactions are disrupted.

\section{Detailed NDCG@10 Results for Query Compression}
\label{app:ndcg10_query_detailed}

Table~\ref{tab:merged-compression-results} reports only the \emph{relative} NDCG@10 changes (\%) from Query Compression. For completeness, Table~\ref{tab:ndcg10_query_detailed_table} and ~\ref{tab:ndcg10_query+doc_detailed_table} provides the corresponding \emph{absolute} NDCG@10 values, with the relative change shown in parentheses.

\begin{table*}[h]
\centering
\renewcommand{\arraystretch}{1.1}
\setlength{\tabcolsep}{1pt}
\resizebox{\textwidth}{!}{%
\begin{tabular}{l ccccccccc}
\toprule
\textbf{Dataset} & \textbf{MiniLM} & \textbf{GTE} & \textbf{Instr.} & \textbf{Dis-Ro} & \textbf{MP-All} & \textbf{MP-QA} & \textbf{Sent-T5} & \textbf{BGE} & \textbf{SFR} \\
\midrule
\textbf{Apps} & \cellcolor{red!25}6.3~(-4.1) & \cellcolor{green!25}58.3~(+1.1) & \cellcolor{red!25}7.8~(-14.5) & \cellcolor{red!25}3.1~(-16.2) & \cellcolor{green!25}8.5~(+1.1) & \cellcolor{red!25}6.8~(-11.1) & \cellcolor{green!25}7.7~(+18.0) & \cellcolor{red!25}12.4~(-16.1) & \cellcolor{red!25}49.5~(-0.2) \\
\textbf{SciDocs} & \cellcolor{red!25}21.6~(-0.2) & \cellcolor{green!25}23.1~(+18.5) & \cellcolor{green!25}19.1~(+3.0) & \cellcolor{red!25}21.6~(-0.5) & \cellcolor{red!25}23.5~(-1.1) & \cellcolor{green!25}18.0~(+3.3) & \cellcolor{green!25}16.6~(+3.8) & \cellcolor{green!25}17.1~(+4.8) & \cellcolor{green!25}25.9~(+2.0) \\
\textbf{MedQA} & \cellcolor{green!25}61.6~(+2.1) & \cellcolor{green!25}72.9~(+2.0) & \cellcolor{green!25}68.8~(+2.1) & \cellcolor{green!25}61.9~(+2.7) & \cellcolor{green!25}68.0~(+2.2) & \cellcolor{green!25}67.7~(+1.9) & \cellcolor{green!25}65.0~(+6.9) & \cellcolor{green!25}68.4~(+0.4) & \cellcolor{green!25}75.1~(+0.5) \\
\textbf{ArguAna} & \cellcolor{green!25}50.8~(+1.2) & \cellcolor{red!25}72.8~(-2.2) & \cellcolor{green!25}57.2~(+3.0) & \cellcolor{green!25}48.2~(+0.6) & \cellcolor{green!25}46.6~(+0.3) & \cellcolor{red!25}50.9~(-0.3) & \cellcolor{green!25}43.8~(+11.1) & \cellcolor{green!25}55.2~(+2.2) & \cellcolor{red!25}69.9~(-1.3) \\
\textbf{StackOverflow} & \cellcolor{green!25}84.0~(+0.1) & \cellcolor{red!25}91.1~(-0.1) & \cellcolor{red!25}84.0~(-1.1) & \cellcolor{green!25}72.1~(+0.3) & \cellcolor{red!25}90.2~(-0.1) & \cellcolor{green!25}82.6~(+1.2) & \cellcolor{green!25}83.8~(+2.9) & \cellcolor{green!25}80.9~(+0.4) & \cellcolor{red!25}89.1~(-0.9) \\
\textbf{TV2Nord} & \cellcolor{green!25}47.8~(+40.2) & \cellcolor{green!25}74.5~(+4.4) & \cellcolor{green!25}75.8~(+2.8) & \cellcolor{green!25}45.8~(+17.1) & \cellcolor{green!25}49.5~(+17.7) & \cellcolor{green!25}56.9~(+7.5) & \cellcolor{green!25}69.1~(+1.2) & \cellcolor{green!25}94.6~(+0.3) & \cellcolor{green!25}72.2~(+5.5) \\
\textbf{GerDa} & \cellcolor{green!25}2.6~(+6.2) & \cellcolor{green!25}15.2~(+3.5) & \cellcolor{green!25}13.0~(+15.7) & \cellcolor{red!25}4.2~(-1.4) & \cellcolor{green!25}4.1~(+7.4) & \cellcolor{green!25}5.5~(+7.0) & \cellcolor{green!25}7.2~(+3.2) & \cellcolor{green!25}32.6~(+1.8) & \cellcolor{green!25}15.4~(+7.8) \\
\textbf{ARC} & \cellcolor{green!25}9.7~(+2.5) & \cellcolor{green!25}13.5~(+1.4) & \cellcolor{green!25}14.1~(+12.9) & \cellcolor{green!25}10.7~(+2.7) & \cellcolor{green!25}12.2~(+3.0) & \cellcolor{green!25}11.1~(+2.0) & \cellcolor{green!25}18.0~(+11.1) & \cellcolor{green!25}9.7~(+7.3) & \cellcolor{green!25}14.2~(+4.4) \\
\textbf{FeedbackQA} & \cellcolor{red!25}57.2~(-1.9) & \cellcolor{green!25}70.6~(+1.0) & \cellcolor{red!25}62.7~(-3.7) & \cellcolor{green!25}48.8~(+0.2) & \cellcolor{red!25}58.1~(-2.2) & \cellcolor{red!25}60.3~(-3.2) & \cellcolor{red!25}62.8~(-2.4) & \cellcolor{red!25}69.0~(-0.4) & \cellcolor{red!25}71.0~(-0.2) \\
\textbf{FaithDial} & \cellcolor{green!25}24.1~(+0.3) & \cellcolor{green!25}24.2~(+5.1) & \cellcolor{green!25}24.0~(+4.0) & \cellcolor{green!25}24.5~(+0.8) & \cellcolor{green!25}24.7~(+0.8) & \cellcolor{green!25}24.9~(+1.8) & \cellcolor{green!25}24.2~(+4.2) & \cellcolor{green!25}21.7~(+7.9) & \cellcolor{green!25}26.2~(+4.8) \\
\textbf{Code} & \cellcolor{green!25}78.4~(+1.2) & \cellcolor{green!25}90.2~(+2.3) & \cellcolor{red!25}90.7~(-1.0) & \cellcolor{green!25}82.4~(+6.0) & \cellcolor{green!25}78.1~(+1.1) & \cellcolor{green!25}81.9~(+1.3) & \cellcolor{green!25}87.3~(+38.7) & \cellcolor{green!25}92.5~(+4.0) & \cellcolor{red!25}97.3~(-0.2) \\
\textbf{MLQA} & \cellcolor{green!25}61.4~(+0.5) & \cellcolor{green!25}70.3~(+0.6) & \cellcolor{green!25}67.7~(+2.5) & \cellcolor{green!25}61.9~(+0.8) & \cellcolor{green!25}62.7~(+1.0) & \cellcolor{green!25}64.6~(+0.4) & \cellcolor{red!25}60.6~(-0.4) & \cellcolor{green!25}70.9~(+0.2) & \cellcolor{green!25}68.8~(+1.8) \\
\textbf{NarrativeQA} & \cellcolor{green!25}20.2~(+11.0) & \cellcolor{green!25}42.6~(+41.5) & \cellcolor{green!25}27.8~(+7.1) & \cellcolor{green!25}18.3~(+15.3) & \cellcolor{green!25}21.8~(+13.1) & \cellcolor{green!25}21.7~(+5.8) & \cellcolor{green!25}24.8~(+20.5) & \cellcolor{green!25}49.5~(+1.6) & \cellcolor{green!25}40.2~(+2.5) \\
\textbf{SpartQA} & \cellcolor{green!25}11.9~(+620.0) & \cellcolor{green!25}19.5~(+149.5) & \cellcolor{green!25}7.0~(+343.0) & \cellcolor{green!25}10.9~(+508.4) & \cellcolor{green!25}4.5~(+1954.5) & \cellcolor{green!25}7.8~(+511.0) & \cellcolor{green!25}4.6~(+119.7) & \cellcolor{green!25}21.9~(+192.5) & \cellcolor{green!25}11.2~(+849.1) \\
\bottomrule
\end{tabular}%
}
\caption{Raw NDCG@10 under Query Compression, with the relative change from the uncompressed baseline shown in parentheses (\%). This table provides the absolute values corresponding to the improvement-only summary in Table~\ref{tab:merged-compression-results}. Positive gains are highlighted \colorbox{green!25}{green}, negative values are \colorbox{red!25}{red}.}
\label{tab:ndcg10_query_detailed_table}
\end{table*}

\begin{table*}[h]
\centering
\scriptsize
\renewcommand{\arraystretch}{1.1}
\setlength{\tabcolsep}{1pt}
\resizebox{\textwidth}{!}{%
\begin{tabular}{l ccccccccc}
\toprule
\multirow{2}{*}{\textbf{Datasets}} & \multicolumn{9}{c}{\textbf{Query+Document Compression}} \\
\cmidrule(lr){2-10}
 & \textbf{MiniLM} & \textbf{GTE} & \textbf{Instr.} & \textbf{Dis-Ro} & \textbf{MP-All} & \textbf{MP-QA} & \textbf{Sent-T5} & \textbf{BGE} & \textbf{SFR} \\
\midrule
\textbf{Apps} & \cellcolor{red!25}5.2~(-21.1) & \cellcolor{red!25}56.8~(-1.5) & \cellcolor{red!25}6.8~(-24.8) & \cellcolor{red!25}1.8~(-52.8) & \cellcolor{red!25}7.4~(-12.4) & \cellcolor{red!25}6.0~(-22.1) & \cellcolor{blue!25}6.9~(+5.0) & \cellcolor{red!25}9.3~(-36.8) & \cellcolor{red!25}49.0~(-1.1) \\
\textbf{SciDocs} & \cellcolor{red!25}21.6~(-0.4) & \cellcolor{blue!25}22.0~(+13.3) & \cellcolor{blue!25}19.0~(+2.5) & \cellcolor{blue!25}21.8~(+0.5) & \cellcolor{red!25}23.7~(-0.3) & \cellcolor{blue!25}17.8~(+2.4) & \cellcolor{red!25}15.1~(-5.2) & \cellcolor{red!25}15.0~(-7.9) & \cellcolor{blue!25}25.8~(+1.9) \\
\textbf{MedQA} & \cellcolor{blue!25}61.5~(+1.9) & \cellcolor{blue!25}72.5~(+1.4) & \cellcolor{blue!25}68.5~(+1.6) & \cellcolor{blue!25}61.6~(+2.2) & \cellcolor{blue!25}67.4~(+1.3) & \cellcolor{blue!25}67.0~(+0.9) & \cellcolor{blue!25}64.2~(+5.6) & \cellcolor{red!25}67.8~(-0.4) & \cellcolor{blue!25}75.0~(+0.5) \\
\textbf{ArguAna} & \cellcolor{blue!25}50.4~(+0.4) & \cellcolor{red!25}72.6~(-2.5) & \cellcolor{blue!25}56.4~(+1.6) & \cellcolor{red!25}47.8~(-0.4) & \cellcolor{red!25}46.0~(-1.1) & \cellcolor{red!25}50.3~(-1.4) & \cellcolor{blue!25}42.2~(+7.1) & \cellcolor{blue!25}54.2~(+0.4) & \cellcolor{red!25}69.4~(-2.0) \\
\textbf{StackOverflow} & \cellcolor{blue!25}84.1~(+0.1) & \cellcolor{blue!25}91.2~(+0.0) & \cellcolor{red!25}84.0~(-1.1) & \cellcolor{red!25}71.3~(-0.8) & \cellcolor{red!25}90.3~(-0.1) & \cellcolor{blue!25}82.0~(+0.5) & \cellcolor{blue!25}83.8~(+2.9) & \cellcolor{red!25}80.5~(-0.1) & \cellcolor{red!25}89.0~(-1.1) \\
\textbf{TV2Nord} & \cellcolor{blue!25}42.6~(+24.9) & \cellcolor{blue!25}73.3~(+2.7) & \cellcolor{blue!25}75.6~(+2.6) & \cellcolor{blue!25}43.7~(+11.7) & \cellcolor{blue!25}43.8~(+4.2) & \cellcolor{blue!25}55.1~(+4.0) & \cellcolor{red!25}67.9~(-0.6) & \cellcolor{red!25}94.1~(-0.2) & \cellcolor{blue!25}70.2~(+2.7) \\
\textbf{GerDa} & \cellcolor{red!25}2.4~(-1.7) & \cellcolor{red!25}13.2~(-10.1) & \cellcolor{blue!25}11.4~(+2.0) & \cellcolor{red!25}3.8~(-10.8) & \cellcolor{red!25}3.7~(-1.6) & \cellcolor{red!25}5.1~(-0.6) & \cellcolor{red!25}6.4~(-7.9) & \cellcolor{red!25}27.5~(-14.2) & \cellcolor{red!25}12.4~(-13.4) \\
\textbf{ARC} & \cellcolor{blue!25}9.8~(+3.7) & \cellcolor{blue!25}13.5~(+1.7) & \cellcolor{blue!25}14.1~(+13.1) & \cellcolor{blue!25}10.6~(+2.3) & \cellcolor{blue!25}12.3~(+3.8) & \cellcolor{blue!25}11.1~(+1.9) & \cellcolor{blue!25}17.7~(+9.4) & \cellcolor{blue!25}9.8~(+8.2) & \cellcolor{blue!25}14.0~(+3.0) \\
\textbf{FeedbackQA} & \cellcolor{red!25}56.7~(-2.7) & \cellcolor{blue!25}70.2~(+0.5) & \cellcolor{red!25}61.2~(-6.0) & \cellcolor{red!25}45.4~(-6.9) & \cellcolor{red!25}57.3~(-3.5) & \cellcolor{red!25}59.0~(-5.3) & \cellcolor{red!25}61.4~(-4.6) & \cellcolor{red!25}68.5~(-1.2) & \cellcolor{red!25}70.3~(-1.2) \\
\textbf{FaithDial} & \cellcolor{blue!25}24.4~(+1.3) & \cellcolor{blue!25}24.2~(+4.8) & \cellcolor{blue!25}23.6~(+2.3) & \cellcolor{red!25}24.3~(-0.2) & \cellcolor{blue!25}24.9~(+1.4) & \cellcolor{blue!25}24.7~(+0.8) & \cellcolor{blue!25}24.0~(+3.1) & \cellcolor{blue!25}21.7~(+7.9) & \cellcolor{blue!25}26.2~(+4.6) \\
\textbf{Code} & \cellcolor{red!25}76.9~(-0.7) & \cellcolor{blue!25}89.0~(+1.0) & \cellcolor{red!25}90.0~(-1.8) & \cellcolor{blue!25}78.3~(+0.7) & \cellcolor{red!25}76.5~(-1.0) & \cellcolor{red!25}79.8~(-1.3) & \cellcolor{blue!25}74.8~(+18.8) & \cellcolor{red!25}87.8~(-1.2) & \cellcolor{red!25}96.9~(-0.6) \\
\textbf{MLQA} & \cellcolor{blue!25}61.2~(+0.2) & \cellcolor{blue!25}70.3~(+0.6) & \cellcolor{blue!25}67.7~(+2.5) & \cellcolor{blue!25}61.6~(+0.4) & \cellcolor{blue!25}62.6~(+0.8) & \cellcolor{blue!25}64.5~(+0.3) & \cellcolor{red!25}59.8~(-1.6) & \cellcolor{blue!25}71.0~(+0.3) & \cellcolor{blue!25}68.7~(+1.7) \\
\textbf{NarrativeQA} & \cellcolor{blue!25}19.5~(+7.1) & \cellcolor{blue!25}41.3~(+37.0) & \cellcolor{blue!25}27.2~(+4.7) & \cellcolor{blue!25}17.7~(+11.6) & \cellcolor{blue!25}20.9~(+8.5) & \cellcolor{blue!25}20.8~(+1.5) & \cellcolor{blue!25}24.1~(+17.4) & \cellcolor{blue!25}49.2~(+1.1) & \cellcolor{blue!25}39.8~(+1.6) \\
\textbf{SpartQA} & \cellcolor{blue!25}12.0~(+626.7) & \cellcolor{blue!25}21.4~(+173.4) & \cellcolor{blue!25}12.9~(+717.7) & \cellcolor{blue!25}10.1~(+463.1) & \cellcolor{blue!25}3.5~(+1509.1) & \cellcolor{blue!25}9.3~(+632.3) & \cellcolor{blue!25}7.2~(+248.1) & \cellcolor{blue!25}23.1~(+208.9) & \cellcolor{blue!25}11.0~(+829.7) \\
\bottomrule
\end{tabular}%
}
\caption{Raw NDCG@10 under Query+Document Compression, with the relative change from the uncompressed baseline shown in parentheses (\%). This table provides the absolute values corresponding to the improvement-only summary in Table~\ref{tab:merged-compression-results}. Positive gains are highlighted \colorbox{blue!25}{blue}, negative values are \colorbox{red!25}{red}.}
\label{tab:ndcg10_query+doc_detailed_table}
\end{table*}

\section{SpartQA}
\label{App:Large_impr}
SpartQA is not a typical retrieval dataset, where in MTEB's setting, the multiple-choice answers for each query are all dumped together, forming a corpus of 1.5k instances. Without PCA, some of the models initially have nearly zero NDCG@10. This is because without knowing that SpartQA focuses on spatial reasoning and that questions needs to be distinguished by the objects and colors occurred, the semantics of texts in SpartQA are highly similar and therefore less discriminative (i.e., all about spatial relations, colors, and shapes).

\section{Failure Analysis}
\label{sec:pca-failure-analysis}

\paragraph{Query Level Effects} 
We present an example where PCA compression reduces the performance. On \textit{FeedbackQA} + \textit{Sent-T5} encoder, the \textsc{NDCG@10} drops by \textbf{2.4\%}.  
Although the loss is small, a closer look reveals a nuanced redistribution of performance rather than a uniform degradation. There are 119 queries improved (the ground-truth document was ranked higher after PCA), while 179 queries regressed (the ground-truth document was ranked lower after PCA). Hence, PCA reshapes the embedding space in a way that facilitates retrieval on a non-trivial subset of queries even while slightly hurting the mean.

\paragraph{Similarity-Score Distributions}

We compared cosine-similarity distributions between queries and their \textit{relevant} (ground truth) versus \textit{non-relevant} (hard-negative) documents before and after PCA (See Table~\ref{tab:fail})

\begin{table}[h]
\resizebox{\columnwidth}{!}{%
\begin{tabular}{lcc}
\toprule
 & \textbf{Original} & \textbf{PCA-compressed} \\
\midrule
\textit{Relevant}     & $\mu{=}0.8361,\;\sigma{=}0.0460$ & $\mu{=}0.4700,\;\sigma{=}0.1402$ \\
\textit{Non-relevant} & $\mu{=}0.6923,\;\sigma{=}0.0546$ & $\mu{=}0.0038,\;\sigma{=}0.1236$ \\
\bottomrule
\end{tabular}%
}
\caption{Cosine Similarity distribution for relevant and non-relevant documents before and after PCA.}
\label{tab:fail}

\end{table}

PCA increases the mean gap between relevant and non-relevant scores (better class separation), but it also inflates the standard deviation for both classes.  
The heavier tails create a larger overlap region near the decision boundary, introducing ambiguity for borderline documents.  
These additional borderline cases explain the modest overall decline in \textsc{NDCG}: PCA resolves some rank-ordering errors but simultaneously introduces new ones. The mixed outcome underscores that PCA’s impact is query-specific: it can sharpen the signal for certain queries while blurring it for others.

\section{Dataset Details}
\label{App:dataset_details}

\begin{table*}[ht]
\centering
\small
\renewcommand{\arraystretch}{1.1}
\setlength{\tabcolsep}{6pt}
\begin{tabular}{lrrll}
\toprule
\textbf{Dataset} & \#Queries & \#Docs & \textbf{Language} & \textbf{Content Category} \\
\midrule
AILACasedocs                    &   50 &   186  & English        & Legal Case Documents          \\
CodeTransOceanDL                &  180 &   816  & Code           & Code Translation          \\
CosQA                           &  500 & 20604  & English        & Code QA                       \\
StackOverflowQA                 & 1994 & 19931  & English        & Programming Q\&A              \\
TV2NordRetrieval                & 2048 &  2048  & Danish         & News Articles                 \\
GerDaLIRSmall                   &12234 &  9969  & German         & Legal Documents               \\
LegalQuAD                       &  200 &   200  & German         & Legal Q\&A                    \\
AILAStatutes                    &   50 &    82  & English        & Legal Statutes                \\
ARCChallenge                    & 1172 &  9350  & English        & Science QA                    \\
BuiltBenchRetrieval             &  334 &  2761  & English        & Engineering Retrieval         \\
FeedbackQARetrieval             & 1992 &  2364  & English        & Public Service QA             \\
MedicalQARetrieval              & 2048 &  2048  & English        & Medical QA                    \\
ChemNQRetrieval                 &   27 & 22933  & English        & Chemistry QA                  \\
AppsRetrieval                   & 3765 &  8765  & English        & Code Snippet Retrieval        \\
FiQA2018                        &  648 & 57638  & English        & Finance QA                    \\
NFCorpus                        &  323 &  3633  & English        & Biomedical Abstracts          \\
SCIDOCS                         & 1000 & 25657  & English        & Scientific Papers             \\
SciFact                         &  300 &  5183  & English        & Scientific Fact Verification  \\
COIRCodeSearchNetRetrieval      &14918 &280310  & Code           & Code Search Retrieval         \\
ArguAna                         & 1406 &  8674  & English        & Argumentative Texts           \\
TRECCOVID                       &   50 &171332  & English        & COVID-19 QA                   \\
MLQARetrieval                   &11582 &  9916  & English        & General QA                    \\
NarrativeQARetrieval            &10557 &   355  & English        & Narrative QA                  \\
SpartQA                         & 3594 &  1592  & English        & Spatial Reasoning QA          \\
FaithDial                       & 2042 &  3539  & English        & Dialogue                       \\
\bottomrule
\end{tabular}
\caption{Overview of retrieval datasets: number of queries and documents, primary language, and content category.}
\label{tab:dataset_summary}
\end{table*}

The 14 datasets from MTEB with sufficient queries are:
\texttt{COIRCodeSearchNetRetrieval} (python subset)~\cite{li2024coir},  \texttt{AppsRetrieval}~\cite{hendrycks2021measuring},  \texttt{ArguAna}~\cite{wachsmuth-etal-2018-retrieval}, \texttt{MedicalQARetrieval}~\cite{benabacha2019question},  \texttt{SciDocs}\cite{cohan-etal-2020-specter}, \texttt{StackOverflowQA}~\cite{li2024coir}, 
\texttt{TV2Nord}~\cite{enevoldsen2024scandinavian}, 
\texttt{ARC}~\cite{clark2018think},
\texttt{GerDa}~\cite{wrzalik-krechel-2021-gerdalir},  \texttt{FeedbackQA}~\cite{li-etal-2022-using}, \texttt{FaithDial}~\cite{dziri-etal-2022-faithdial}, \texttt{NarrativeQA}~\cite{kocisky-etal-2018-narrativeqa}, \texttt{MLQA} (eng-eng subset)~\cite{lewis-etal-2020-mlqa},
and SpartQA~\cite{mirzaee-etal-2021-spartqa}.

The 11 datasets with not enough queries are:
\texttt{SciFact}~\cite{wadden-etal-2020-fact}, 
\texttt{ChemNQRetrieval}~\cite{kasmaee2024chemteb}, \texttt{FiQA2018}~\cite{yang2018financial}, \texttt{NFCorpus}~\cite{boteva2016full},
\texttt{TRECCOVID}~\cite{treccovid}, 
\texttt{Legal}~\cite{9723721}, 
\texttt{CodeTransDL}~\cite{yan-etal-2023-codetransocean},
\texttt{CosQA}~\cite{huang-etal-2021-cosqa}, \texttt{BuiltBench}~\cite{shahinmoghadam2024benchmarking}, \texttt{AILAStat} and \texttt{AILADocs}~\cite{AILA}. 

More details on each dataset has been given in Table~\ref{tab:dataset_summary}.

% \label{app:models}
% \begin{table}[h!]
% \centering
% \renewcommand{\arraystretch}{1.1}
% \begin{adjustbox}{max width=0.6\textwidth}
% \begin{tabular}{@{}lll@{}}
% \toprule
% \textbf{Abbreviation} & \textbf{Full Model Name} & \textbf{Reference} \\
% \midrule
% \texttt{Dis-Ro} & \texttt{all-distilroberta-v1} & \multirow{5}{*}{\cite{reimers-gurevych-2019-sentence}} \\
% \texttt{MP-QA} & \texttt{multi-qa-mpnet-base-dot-v1} & \\
% \texttt{MiniLM} & \texttt{all-MiniLM-L6-v2} & \\
% \texttt{MP-All} & \texttt{all-mpnet-base-v2} & \\
% \texttt{Sent-T5} & \texttt{sentence-t5-xl} & \\
% \midrule
% \texttt{BGE} & \texttt{BGE\_M3} & \cite{chen-etal-2024-m3} \\
% \texttt{GTE} & \texttt{GTE} & \cite{zhang-etal-2024-mgte} \\
% \texttt{SFR} & \texttt{SFRCode} & \cite{liu2024codexembed} \\
% \texttt{Instr.} & \texttt{Instructor} & \cite{su-etal-2023-one} \\
% \bottomrule
% \end{tabular}
% \end{adjustbox}

% \caption{Details of the pretrained dense retrieval models evaluated in this work.}
% \label{tab:model_details}
% \end{table}
\section{Model Details}
\label{app:models}

\begin{table}[h]
\centering
\small
\setlength{\tabcolsep}{6pt}
\renewcommand{\arraystretch}{1.12}

\begin{tabularx}{\linewidth}{@{}lX@{}}
\toprule
\textbf{Abbrev.} & \textbf{Full model name} \\
\midrule

\multicolumn{2}{@{}l}{\emph{Sentence-Transformers family} \citep{reimers-gurevych-2019-sentence}} \\
\texttt{Dis-Ro} & \texttt{all-distilroberta-v1} \\
\texttt{MP-QA} & \texttt{multi-qa-mpnet-base-dot-v1} \\
\texttt{MiniLM} & \texttt{all-MiniLM-L6-v2} \\
\texttt{MP-All} & \texttt{all-mpnet-base-v2} \\

\addlinespace[2pt]
\multicolumn{2}{@{}l}{\emph{Other embedding models}} \\
\texttt{Sent-T5} & \texttt{sentence-t5-xl} \citep{ni-etal-2022-sentence} \\
\texttt{BGE} & \texttt{BGE\_M3} \citep{chen-etal-2024-m3} \\
\texttt{GTE} & \texttt{GTE} \citep{zhang-etal-2024-mgte} \\
\texttt{SFR} & \texttt{SFRCode} \citep{liu2024codexembed} \\
\texttt{Instr.} & \texttt{Instructor} \citep{su-etal-2023-one} \\

\bottomrule
\end{tabularx}
\caption{Pretrained dense retrieval models evaluated in this work.}
\label{tab:model_details}
\end{table}

Table~\ref{tab:model_details} lists the pretrained dense retrieval models used in our evaluation, along with the abbreviations used throughout the paper.

\end{document}